\documentclass[a4paper,11pt]{article}
\usepackage{jheppub,xcolor} 
\usepackage{float}
\usepackage{graphicx,braket,dsfont}
\usepackage[normalem]{ulem}
\usepackage{tikz}
\usetikzlibrary{arrows.meta,calc}
 \usetikzlibrary{arrows.meta,calc,decorations.pathmorphing}
 \usetikzlibrary{
  calc,
  arrows.meta,
  positioning,
  decorations.pathmorphing,
  shadows.blur
}
\usepackage{orcidlink}

\title{\boldmath Integrability breaking in semiclassical strings in Koopman-Krylov space}

\author[a,b,c]{Rathindra Nath Das,}
\author[d]{Saskia Demulder}
\affiliation[a]{Department of Particle Physics and Astrophysics, Weizmann Institute of Science,\\
Rehovot 7610001, Israel}
\affiliation[b]{Center for Theoretical Physics — a Leinweber Institute, MIT, Cambridge, MA 02139, USA}
\affiliation[c]{Institute for Theoretical Physics and Astrophysics, and\\ W\"urzburg-Dresden Cluster of Excellence ctd.qmat,\\
Julius-Maximilians-Universit\"at W\"urzburg, Am Hubland, 97074 W\"urzburg, Germany}
\affiliation[d]{Department of Quantitative Methods, CUNEF Universidad,\\ Calle Almansa 101,
28040 Madrid, Spain}

 \emailAdd{rathindra-nath.das@weizmann.ac.il, saskia.demulder@cunef.edu}

\abstract{
While very powerful, integrability in semiclassical string solutions is known to be a rare property. Motivated by the need to understand and characterise the large landscape of non-integrable string dynamics, we extend Krylov methods for probing chaos to classical systems. We introduce a Koopman-Krylov framework, formulated in the Koopman-von Neumann description of classical mechanics and implemented via a generator
extended dynamic mode decomposition (gEDMD) approximation of the Koopman generator acting on observables. Using this framework, we study how integrability-breaking deformations of integrable string dynamics induce characteristic redistributions of spectral weight, leading to observable-dependent delocalisation and spreading in Krylov space. We illustrate the Koopman-Krylov diagnostics across three classes of non-integrable semiclassical string solutions.}

\begin{document}
\maketitle
\flushbottom

\section{Introduction and motivation}

Non-integrability and chaos are generic in nonlinear dynamics, and integrable systems should be viewed as exceptional. A canonical illustration is provided by the kicked rotor, where small perturbations destroy invariant tori in resonant regions and produce thin chaotic layers and long-lived trapping near remnants of regular structures. String theory does not evade this principle \cite{Stepanchuk:2012xi,Basu:2011fw}. On the contrary, classical integrability in the planar AdS/CFT correspondence has played a central guiding role in uncovering exact structures of the duality, yet it has long been understood that this integrable sector is fragile. Small deformations driven, for example, by higher-loop corrections to the dilatation operator of $\mathcal{N}=4$ SYM\footnote{While higher-loop corrections to the integrable one-loop dilatation operator \cite{Minahan:2002ve,Beisert:2003tq} break integrability, substantial evidence suggests that the all-order planar dilatation operator is again integrable \cite{Beisert:2005fw,Beisert:2006ez}. An integrable long-range spin chain believed to capture the asymptotic all-loop planar spectrum was constructed in \cite{Bargheer:2008jt}.} \cite{McLoughlin:2022jyt} or by changes of the compact part of the string background generically induce non-integrable dynamics.  Importantly, as emphasised in \cite{Giataganas:2014hma}, non-integrability does not necessarily imply fully developed chaotic behaviour, a distinction that will play a central role in our analysis.

The breakdown of integrability in semiclassical string reductions has primarily been investigated using standard tools from nonlinear dynamics applied to finite-dimensional Hamiltonian truncations. When a string ansatz reduces the worldsheet dynamics to an effective phase space with a conserved Hamiltonian, integrable motion organises trajectories into quasi-periodic invariant tori on fixed-energy surfaces. The onset of non-integrability is then visible in Poincar\'e sections through the erosion of regular invariant curves predicted by Kolmogorov-Arnold-Moser (KAM) theory and the emergence of stochastic layers signalling local chaotic transport and long-lived trapping near remnants of regular islands. Complementary evidence is provided by Lyapunov exponents, which quantify sensitivity to initial conditions and distinguish chaotic trajectories from long transients. These tools have been implemented in a wide range of semiclassical string models, including winding-string dynamics on $\mathrm{AdS}_5\times T^{1,1}$, where the progressive erosion of invariant structures demonstrates that non-integrable behaviour arises on open sets of initial data rather than as a fine-tuned exception \cite{PandoZayas:2010xpn,Basu:2016zkr,Panigrahi:2016zny,Giataganas:2017guj,Cubrovic:2019qee,Akutagawa:2019awh,Ishii:2021asw,McLoughlin:2022jyt,Djukic:2023dgk,Chakraborty:2025crb,Berenstein:2025ese}. Analytic approaches based on differential Galois theory and Kovacic's theorem provide complementary obstructions to integrability by establishing the absence of additional meromorphic first integrals \cite{Basu:2011fw,Stepanchuk:2012xi,Chervonyi:2013eja,Giataganas:2014hma,Asano:2016qsv,Giataganas:2017guj,Frolov:2017kze,Nunez:2018qcj,Filippas:2019ihy,Giataganas:2019xdj,Rigatos:2020igd,Rigatos:2020hlq,Nayak:2025kbl}.

While these approaches provide powerful criteria for detecting non-integrability, they do not offer a systematic probe of how integrable motion begins to fail. In near-integrable regimes the phase space is typically mixed; with large regular regions coexisting with thin chaotic layers and resonant transport channels. In such situations, departures from integrable dynamics need not manifest as uniform exponential instability, but rather through gradual redistribution of dynamical structures and mixing between specific directions in phase space. This motivates probes that resolve how observables evolve under the flow, rather than focusing solely on the divergence of individual trajectories.

In this work we adopt such an observable-based perspective by studying the spreading of classical observables under Koopman evolution. The Koopman-von Neumann \cite{koopman1931hamiltonian,neumann1932operatorenmethode} formulation of classical dynamics describes the Hamiltonian flow as a linear evolution acting on functions on phase space, generated by the Liouville operator $\mathcal L=\{\cdot,H\}$. Repeated action of $\mathcal L$ generates Krylov subspaces spanned by $\{g,\mathcal L g,\mathcal L^2 g,\ldots\}$, providing a classical analogue of operator growth familiar from quantum many-body systems. Inspired by recent developments in quantum Krylov complexity and operator spreading 
\cite{Parker:2018yvk,Caputa:2021sib,Balasubramanian:2022tpr}, we formulate an  observable-based Koopman-Krylov framework tailored to classical Hamiltonian dynamics.  Unlike state- or operator-Krylov constructions in quantum systems, which rely on  unitary evolution in a quantum Hilbert space of states or operators, our construction acts on  classical observables through Koopman evolution and is therefore intrinsically suited  to nonlinear classical flows.  We approximate the Koopman generator using generator extended dynamic mode decomposition (gEDMD) \cite{Williams2015,klus2020data} and extract finite-resolution information about this spectral measure via Lanczos tridiagonalisation in Krylov space. The guiding principle is that weak classical chaos in semiclassical strings is neither uniform in phase space nor uniform over the classical Hilbert space of observables: different observables couple to different resonant channels and therefore act as distinct probes of incipient integrability breaking.

We apply this Koopman-Krylov framework to study integrability breaking in concrete semiclassical string reductions. Starting from integrable limits of planar AdS/CFT sectors, we analyse weak deformations arising from higher-loop spin-chain corrections and geometric deformations of the target space. We focus on representative examples including deformed $SU(2)$ and $SU(3)$ spin-chain sectors and near-Penrose limits of strings on $\mathrm{AdS}_5\times T^{1,1}$. For these systems we compute observable-resolved Krylov diagnostics and spectral measures to track how integrability is weakened. Unlike quantum chaotic systems, where integrability breaking often produces universal spectral behaviour, we find that classical string dynamics exhibits deformation-dependent signatures: the same observable-based diagnostic probes distinct dynamical channels depending on the underlying mechanism. Koopman-Krylov methods therefore complement Lyapunov and Poincar\'e diagnostics by resolving how observable dynamics and spectral transport reorganise, identifying which dynamical directions begin to mix and how spectral content is redistributed under different integrability-breaking mechanisms.

\noindent
This paper is organised as follows. In section \ref{sec:koopman_complexity} we introduce the observable-based Koopman-Krylov framework for semiclassical string dynamics, including the Koopman spectral measures induced by seed observables and the finite-resolution approximation pipeline based on gEDMD and Lanczos tridiagonalisation. We then define the set of Krylov and spectral diagnostics used throughout the paper, including Krylov spread, delocalisation measures, sector-resolved leakage, and Wasserstein spectral transport. In section \ref{sec:KK_obs} we apply these probes to three representative mechanisms of integrability breaking in semiclassical string solutions: higher-loop corrections in the $SU(2)$ sector, Leigh-Strassler deformations of the $SU(3)$ sector, and near-Penrose limits of $\mathrm{AdS}_5 \times T^{1,1}$ strings. We conclude in section \ref{sec:conclusions} with a discussion and future perspectives. Additional  perturbative analyses and technical comments are collected in the appendices.

\section{Classical integrability breaking and Koopman theory}\label{sec:string_reductions}

We first briefly recall how semiclassical string solutions give rise to finite-dimensional classical Hamiltonian systems. 
The starting point is the bosonic non-linear sigma model describing strings propagating on a fixed target-space background with metric $G_{\mu\nu}(X)$,
\begin{equation}
    S = \frac{\sqrt{\lambda}}{4\pi}\int \mathrm d\tau\, \mathrm d\sigma \, G_{\mu\nu}(X)\,\partial_a X^\mu \partial^a X^\nu\,,
\end{equation}
supplemented by the Virasoro constraints. This defines a two-dimensional field theory with infinitely many degrees of freedom.

To obtain a tractable dynamical system one imposes a symmetry-reduction ansatz that restricts the embedding coordinates to a finite set of collective variables.  Typically one chooses an ansatz in which some target-space coordinates are frozen or rotate with fixed frequencies, while the nontrivial degrees of freedom depend only on a single worldsheet coordinate.  Schematically,
\begin{equation}
X^\mu(\tau,\sigma)\;\longrightarrow\; X^\mu\big(q_i(\sigma)\big)\,,
\end{equation}
and the dynamical variables $q_i(\sigma)$ describe a finite number of collective modes. Substituting this ansatz into the worldsheet action and integrating over the spectator coordinate produces an effective one-dimensional action
\begin{equation}
S_{\rm eff}=\int\mathrm  d\sigma\, L_{\rm eff}(q_i,q_i')\,, \qquad q_i'=\frac{\mathrm dq_i}{\mathrm d\sigma}\,,
\end{equation}
from which one defines canonical momenta $p_i=\frac{\partial L_{\rm eff}}{\partial q_i'}$ and the resulting effective Hamiltonian $H_{\rm eff}(q_i,p_i)=p_i q_i'-L_{\rm eff}$. 

A key feature common to the systems considered in this work is that the nontrivial dynamics is organised along the spatial worldsheet coordinate $\sigma$, rather than along target-space time. After fixing static gauge and imposing the reduction ansatz, the equations of motion take the Hamiltonian form
\begin{equation}
\frac{\mathrm d q_i}{\mathrm d\sigma} = \frac{\partial H_{\rm eff}}{\partial p_i}\,, 
\qquad  \frac{\mathrm d p_i}{\mathrm d\sigma} = -\frac{\partial H_{\rm eff}}{\partial q_i}\,,
\end{equation}
where $(q_i,p_i)$ parametrise the reduced phase space. Time dependence, when present, enters only through fixed frequencies and conserved charges, so in the systems studied here that instability or non-integrable behaviour manifests through mixing along the $\sigma$-flow.

\subsection{Koopman evolution and observable spectral measures}\label{sec:koopman}

\begin{figure}[t]
    \centering
    \includegraphics[width=0.75\linewidth]{ 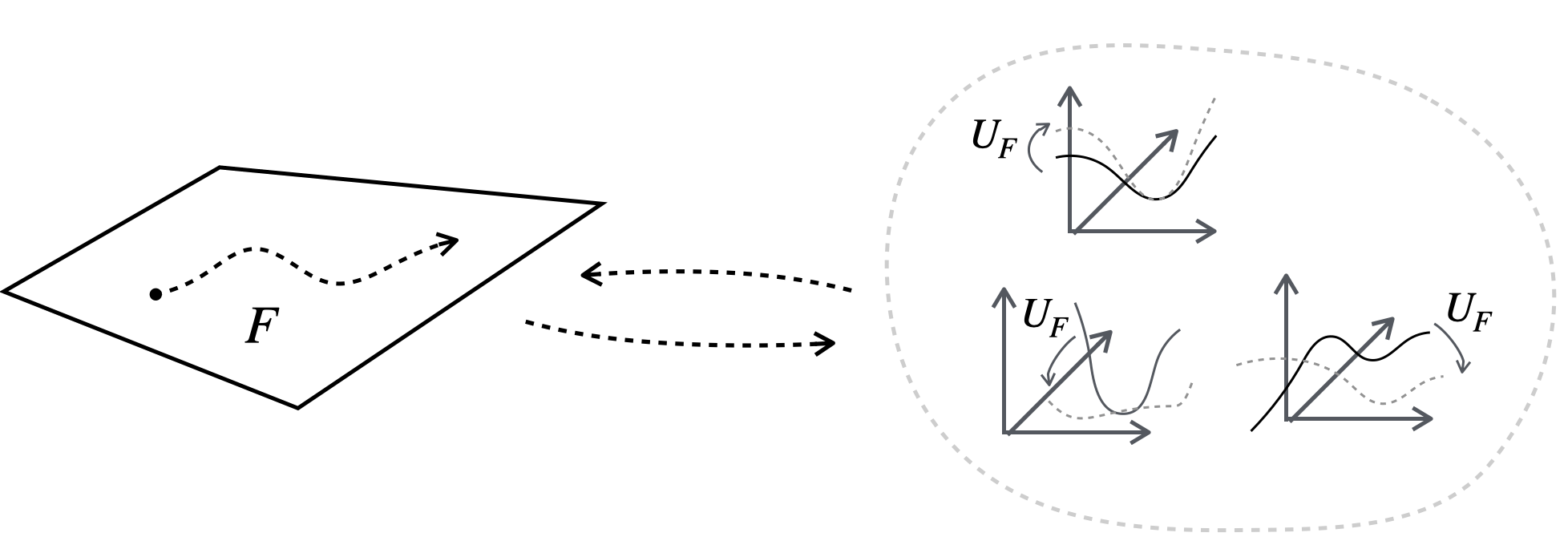}
    \caption{Schematic illustration of the Koopman picture. A nonlinear trajectory generated by a classical flow $F$ on phase space (left) is lifted to linear evolution of observables under the Koopman operator $U_F$ (right), which acts on functions rather than points. Different observables evolve linearly in an common Hilbert space, encoding the same underlying nonlinear dynamics.}
    \label{fig:koopman}
\end{figure}

The Koopman-von Neumann formulation of classical mechanics \cite{koopman1931hamiltonian,neumann1932operatorenmethode} provides a linear representation of nonlinear phase-space dynamics at the level of observables as depicted in fig.~\ref{fig:koopman}. Rather than evolving phase-space points $x(\sigma)\in X$, one studies the induced evolution of observables $g:X\to\mathbb C$ under the deterministic flow $F^\sigma:X\to X$ generated by the reduced Hamiltonian $H_{\rm eff}$. This defines the Koopman operator acting by composition,
\begin{equation}\label{eq:nonlin_flow}
(U^\sigma g)(x)=g(F^\sigma x)\,,
\end{equation}
which is linear even when the underlying flow is nonlinear. We refer to \cite{budivsic2012applied,Williams2015,mezic2005spectral,mezic2022numerical} for detailed reviews on Koopman theory.

Koopman dynamics naturally equips classical mechanics with a Hilbert-space structure. For a choice of invariant measure $\mu$ on $X$ (in our applications the Liouville measure restricted to a narrow energy shell), the space of square-integrable observables
\begin{equation}
\mathcal H_{\rm obs}:=L^2(X,\mu)
\end{equation}
forms a Hilbert space with inner product
\begin{equation}\label{eq:inner-product}
\langle f,g\rangle=\int_X f(x)^*g(x)\,\mathrm d\mu(x)\,.
\end{equation}
If the flow preserves $\mu$, then $U^\sigma$ is unitary on $\mathcal H_{\rm obs}$. The infinitesimal generator $A$ of the Koopman group is defined by
\begin{equation}\label{eq:Koopmanrel}
\frac{d}{d\sigma}U^\sigma f=\mathcal L\,U^\sigma f\,,
\end{equation}
and we introduce the self-adjoint Koopman generator
\begin{equation}\label{eq:Koopmanop}
K:=-i\mathcal L\,, \qquad U^\sigma=e^{-i\sigma K}\,.
\end{equation}
For Hamiltonian flow one has
\begin{equation}
K=-i\{\cdot,H_{\rm eff}\}\,,
\end{equation}
so that classical nonlinear dynamics is lifted to a linear but infinite-dimensional evolution problem on $\mathcal H_{\rm obs}$ with Liouville or Koopman operator $\mathcal L=iK$.  Finally, for a chosen observable $g\in L^2(X,\mu)$, Koopman evolution defines the autocorrelator
\begin{equation}\label{eq:autocorrelator}
C_g(\sigma)=\langle g,U^\sigma f\rangle =\int_{\mathbb R}e^{-i\sigma\omega}\,\mathrm d\mu_g(\omega)\,,
\end{equation}
where $\mu_g$ is the positive spectral measure associated with the pair $(K,g)$. This measure characterises the dynamical content of the observable and depends explicitly on the choice of observable $g$: different observables probe different aspects of the same flow.

Koopman theory therefore recasts classical nonlinear dynamics as a linear evolution problem for observables. Of course, the non-linear nature of the particular differential equations governing the phase space dynamics cannot be reduced to simple linear dynamics, and the Koopman is a linear but infinite dimensional operator. This formulation provides the natural bridge to apply Krylov methods and study how dynamical structures reorganise under the flow through the spectral properties of observables. Before introducing this classical version of Krylov diagnostics for classical dynamical systems, we first review characteristic properties of non-integrable and weakly chaotic classical systems.

\subsection{Weak non-integrability, resonances, and Koopman spectral structure}
\label{sec:weak_chaos_koopman}

The aim of this section is to provide a brief and panoramic overview of some mechanisms and characteristics of integrability breaking deformations, as e.g. realised in the semiclassical string systems studied in this work. In these systems, under any deformation, one never obtains fully chaotic or globally mixing dynamics. Instead, these deformations lead to weakly non-integrable deformations of integrable systems. In this regime, instability is not uniform across phase space but is organised by resonant regions where averaging breaks down: most invariant tori persist, while slow transport and dephasing occur along resonant channels.

\subsubsection{Mechanism of weak integrability breaking}\label{sec:kam}
The Hamiltonian reductions we will consider here will consist of an integrable Hamiltonian together with a perturbatively small integrability breaking term.
More precisely, we begin from an integrable Hamiltonian system, thus admitting action-angle variables,
\begin{equation}
H(I,\theta)=H_0(I)\,,\qquad (I,\theta)\in\mathbb{R}^d\times\mathbb{T}^d\,,
\end{equation}
for which $\dot I=0$ and $\dot\theta=\omega(I)\equiv\partial_I H_0$. In these coordinates, the phase space of the integrable system is locally foliated by invariant tori $I=\mathrm{const.}$ supporting quasi-periodic motion. 

Introducing a weak deformation introduces angle dependence,
\begin{equation}\label{eq:def_Ham}
H_{\varepsilon}(I,\theta)=H_0(I)+\varepsilon H_1(I,\theta)\,,\qquad 0<\varepsilon\ll1\,,
\end{equation}
leads to slow evolution of the actions,
\begin{equation}
\dot I=-\varepsilon\,\partial_\theta H_1(I,\theta)\,.
\end{equation}
Fourier expanding the intregrability-breaking deformation $H_1$,
\begin{equation}
H_1(I,\theta)=\sum_{k\in\mathbb{Z}^d}h_k(I)e^{ik\cdot\theta}\,,
\end{equation}
and inserting the unperturbed motion $\theta(\sigma)=\theta_0+\omega(I)t$ yields the characteristic small-denominator structure
\begin{equation}\label{eq:small_denom}
\Delta I(\sigma) \sim \varepsilon\sum_{k\in\mathbb{Z}^d}k\,h_k(I)\, \frac{e^{i(k\cdot\omega)\sigma}-1}{k\cdot\omega}\,.
\end{equation}
Away from resonances ($|k\cdot\omega|\gtrsim 1$) the response is bounded and oscillatory, while near resonances ($k\cdot\omega\simeq 0$) it is parametrically enhanced. KAM theory formalises that weak breaking of integrability is organised by resonant Fourier channels rather than by uniform instability: quasi-regular regions retain their integrable structure, while thin resonant layers can lead to chaotic motion and slow transport. See e.g. \cite{de2001tutorial,lichtenberg2013regular} for further details and formal treatements on KAM theory.

\subsubsection{Resonant dephasing and finite-resolution packetisation}\label{sec:packetisation}

From the perspective of observables, the distinction between regular motion and resonance-driven transport is reflected not in individual trajectories, but in how spectral content is reorganised under the flow \cite{mezic2005spectral,mezic2013analysis}. This structure is naturally described in the Koopman formulation of sec.~\ref{sec:koopman}. For Hamiltonian flow for the deformed Hamiltonian $H_{\varepsilon}$ in eq.~\eqref{eq:def_Ham}, one constructs the Koopman operator \eqref{eq:Koopmanop}, i.e. $K=-iA=-i\{\cdot,H_{\varepsilon}\}$. One must then distinguish: (i) the spectral type of $K$ on $L^2(X,\mu)$, (ii) the spectral measure $\mu_g$ for a chosen observable $g$ through its autocorrelator \eqref{eq:autocorrelator}, and (iii) the finite-resolution spectra obtained after truncation (necessarily discrete). Crucially, even for a fixed dynamical system, the spectral measure $\mu_g$ depends on the choice of observable $g$, and different observables can probe distinct resonant structures of the same flow.

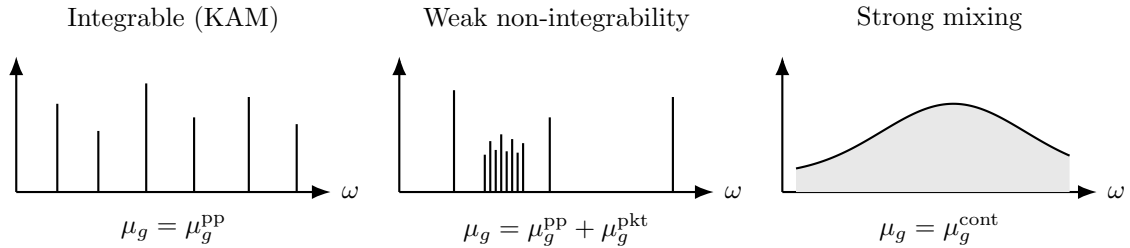
\begin{figure}[t]
\centering
\begin{tikzpicture}[scale=0.9]
\def\dx{5.6}
\def\W{4.6}
\def\H{2.0}

\tikzset{
  axis/.style={thick,-{Latex[length=2.5mm]}},
  spike/.style={thick},
  lbl/.style={font=\small}
}

\node[lbl] at (0.5*\W,\H+0.55) {Integrable (KAM)};
\node[lbl] at (\dx+0.5*\W,\H+0.55) {Weak non-integrability};
\node[lbl] at (2*\dx+0.5*\W,\H+0.55) {Strong mixing};

\foreach \k in {0,1,2} {
  \begin{scope}[shift={(\k*\dx,0)}]
    \draw[axis] (0,0) -- (\W,0) node[right, font=\small] {$\omega$};
    \draw[axis] (0,0) -- (0,\H);
  \end{scope}
}

\begin{scope}[shift={(0,0)}]
  \foreach \w/\h in {0.6/1.3,1.2/0.9,1.9/1.6,2.6/1.1,3.4/1.4,4.1/1.0} {
    \draw[spike] (\w,0) -- (\w,\h);
  }
  \node[lbl] at (0.5*\W,-0.55) {$\mu_g=\mu_g^{\rm pp}$};
\end{scope}

\begin{scope}[shift={(\dx,0)}]
  \foreach \w/\h in {0.8/1.5,2.2/1.1,4.0/1.4} {
    \draw[spike] (\w,0) -- (\w,\h);
  }
  \foreach \w/\h in {1.25/0.55,1.33/0.75,1.41/0.62,1.49/0.85,1.57/0.60,1.65/0.78,1.73/0.58,1.81/0.72} {
    \draw[spike] (\w,0) -- (\w,\h);
  }
  \node[lbl] at (0.5*\W,-0.55) {$\mu_g=\mu_g^{\rm pp}+\mu_g^{\rm pkt}$};
\end{scope}

\begin{scope}[shift={(2*\dx,0)}]
  \path[fill=gray!18, draw=none]
    (0.2,0)
    -- plot[domain=0.2:4.2, samples=90] (\x,{0.25+1.05*exp(-0.45*(\x-2.5)^2)})
    -- (4.2,0) -- cycle;
  \draw[thick]
    plot[domain=0.2:4.2, samples=90] (\x,{0.25+1.05*exp(-0.45*(\x-2.5)^2)});
  \node[lbl] at (0.5*\W,-0.55) {$\mu_g=\mu_g^{\rm cont}$};
\end{scope}

\end{tikzpicture}
\caption{Schematic observable spectral measures under Hamiltonian flow. Integrable dynamics yields pure point measures. Weak non-integrability produces mixed measures, whose finite-resolution signature is ``packetisation'' of selected spectral lines. Fully mixing dynamics corresponds to purely continuous measures.}
\label{fig:KAM_spectral_measures}
\end{figure}

In order to introduce the different spectram types, remember that for an integrable Hamiltonian $H_0(I)$ one has
\begin{equation}
K_{H_0}=-i\{\cdot,H_0\}=-i\sum_i\omega_i(I)\,\partial_{\theta_i}\,,
\end{equation}
with Fourier eigenfunctions $e^{ik\cdot\theta}$ and eigenvalues $k\cdot\omega(I)$ on a fixed invariant torus.  Accordingly, for a smooth observable restricted to such a torus, $\mu_g$ is discrete or pure point.  

In fact, for more general, non-integrable systems it is useful to distinguish three qualitative spectral regimes\footnote{Note that while this is accepted as the general rule, one can find exceptions. In particular in \cite{mezic2020spectrum}, the author discusses a simple example of a non-dissipative system with non-chaotic dynamics which displays a continuous spectrum.} (see fig.~\ref{fig:KAM_spectral_measures}). 
In an integrable regime, the Koopman spectral measure associated with a smooth observable $g$ is purely discrete,
\begin{equation}
    \mu_g = \mu_g^{\rm pp}\,,
\end{equation}
reflecting quasi-periodic motion on invariant tori. In this case autocorrelators are exactly or asymptotically quasi-periodic, and Krylov evolution remains structured even at long times. Any observed spreading arises from kinematic superposition of incommensurate frequencies rather than from dynamical mixing. 

In a near-integrable or weakly non-integrable regime, the spectral measure develops an additional ``packetised'' component $\mu_g^{\rm pkt}$,
\begin{equation}
    \mu_g = \mu_g^{\rm pp} + \mu_g^{\rm pkt}, \qquad \|\mu_g^{\rm pkt}\| > 0\,,
\end{equation}
while sharp spectral lines persist. This situation is characteristic of weak integrability breaking: most invariant tori survive, but resonant layers induce slow dephasing and a redistribution of spectral weight for observables that couple to them. At finite resolution the continuous contribution appears not as a smooth background but as dense clusters of spectral lines, which we refer to as ``packetisation''. The resulting spectral transport, and therefore the mechanism of packet formation, depends on the chosen observable $g$.

Finally, in a mixing regime the Koopman spectrum on the sampled invariant set is continuous for all non-constant observables, so that $\mu_g$ is purely continuous. This corresponds to genuinely mixing dynamics: correlations decay, isolated spectral lines are absent, and Krylov evolution exhibits strong delocalisation largely independent of the chosen observable\footnote{The above classification refers to analytic expressions or ideal infinite-resolution spectral measures. Numerically one always works with finite-dimensional approximations, so even a mixed spectrum is represented by a discrete set of spectral values.}.

Near-integrable Hamiltonian systems typically exhibit coexistence of regular islands and chaotic layers, and the spectral type sampled by a given observable depends on which regions of phase space are explored and how the averaging is performed. Weak non-integrability therefore corresponds neither to pure quasi-periodicity nor to global mixing, but to resonance-organised redistribution of spectral weight for generic observables. In the systems studied below we will in fact observe that spectral transport can be strongly energy-dependent: as the energy increases, additional resonances become accessible, leading to progressively more pronounced packetisation.

\section{Koopman-Krylov framework for classical phase space dynamics}\label{sec:koopman_complexity}

Having recast classical dynamics as linear Koopman evolution on the observable Hilbert space $\mathcal H_{\rm obs}$ in section~\ref{sec:koopman}, we now introduce a Krylov framework for analysing this evolution. Krylov methods characterise linear dynamics by the spreading of a seed vector under repeated action of the generator. In quantum many-body systems this probes operator growth and quantum chaos. Here we apply the same construction to classical dynamics by using the Koopman generator $K$ and classical observables as seeds. Since the Koopman operator $K$ in eq.~\eqref{eq:Koopmanop} is a first-order differential operator and weakly non-integrable systems exhibit strongly observable-dependent dynamics, Krylov spreading in this setting should not be interpreted as a universal chaos diagnostic. Instead it provides an observable-resolved probe of spectral redistribution and transport mechanisms associated with integrability breaking.

\subsection{Koopman-Krylov space and diagnostics}

As reviewed in section \ref{sec:koopman}, Koopman's key observation was that a  nonlinear flow on phase space can be represented as a linear (though generically, infinite-dimensional) evolution on a Hilbert space of observables, by shifting the dynamics from trajectories to functions transported by composition with the flow, see eq.~\eqref{eq:Koopmanrel}. Thus, having reduced the evolution to a linear operator we can therefore define Krylov spreading for observables by replacing the quantum generator with the classical Koopman operator $K$ and the initial state or operator with an observable $g_0\in\mathcal{H}_{\mathrm{obs}}$. The associated Krylov chain is
\begin{equation}
 g_0\,,\quad g_1 = K g_0\,,\quad g_2 = K g_1\,,\quad \dots
\end{equation}
and after Lanczos orthogonalisation one obtains a basis $\{|n)\}$ such that
\begin{equation}
 K|n) = a_n |n) + b_{n+1}|n{+}1) + b_n |n{-}1)\,.
\end{equation}
The evolved observable $g(\sigma)=U^\sigma g_0$, as determined by the non-linear evolution in eq.~\eqref{eq:nonlin_flow}, decomposes as
\begin{equation}
 g(\sigma) = \sum_{n\ge 0} \phi_n(\sigma)\,|n)\,,
\end{equation}
and the Krylov wavefunctions $\phi_n(\sigma)$ quantifies spreading in Koopman-Krylov space. All Krylov diagnostics considered below  are functionals of the spectral measure $\mu_{g_0}$ associated with $(K,g_0)$.

The central difference from Heisenberg or Schr\"odinger evolution in quantum systems is that the Koopman operator $K$ is not a finite dimensional matrix. In order to perform numerical computations, one therefore restricts to a finite approximation space
\begin{equation}\label{eq:dict_DN}
 \mathcal{D}_N = \mathrm{span}\{\phi_1,\dots,\phi_N\} \subset \mathcal{H}_{\mathrm{obs}}\,,
\end{equation}
which induces the compression
\begin{equation}
 K_N = \Pi_N K \Pi_N\,, \quad g_{0,N} = \Pi_N g_0\,.
\end{equation}
Since the pair $(K_N,g_{0,N})$ is now a finite dimensional approximation of the non-linear evolution \eqref{eq:Koopmanrel}, one can apply the Lanczos algorithm as described above and study the associated Koopman-Krylov spread in the observable Hilbert space.

Finally, let us emphasise that our use of Koopman-Krylov methods differs from the conventional objectives of Koopman analysis. Standard Koopman theory and its numerical implementations typically aim to extract spectral data of the evolution operator characterising the evolution flow. Related Krylov-based approaches in the Koopman literature, see e.g. \cite{arbabi2017ergodic,mezic2022numerical}, are likewise designed to approximate Koopman eigensystems or spectral decompositions. Koopman methods have also been employed in the literature as tools to characterise and detect chaotic dynamics through spectral properties of the evolution operator, see e.g. \cite{brunton2017chaos,mezic2013analysis}. In contrast, our goal is not to reconstruct the Koopman spectrum or identify coherent
eigenmodes. Instead, we use Krylov constructions as diagnostics of observable dynamics, probing how spectral weight associated with a seed observable redistributes under the flow and how this redistribution reflects weak integrability breaking. The Krylov data are therefore interpreted as yielding finite-resolution probes of dynamical transport and mixing rather than as approximations to Koopman eigenstructures.

\subsubsection{Krylov diagnostics for semiclassical systems}\label{sec:Krylovtoolkit}

As discussed earlier, in near-integrable regimes the presence of chaos is often diagnosed through Lyapunov exponents or Poincar\'e sections. Such quantities however do not always provide a complete picture of weak integrability breaking. In particular, signatures of weak integrability breaking instead also appear in finer properties of how spectral weight is redistributed in the observable-induced Koopman spectrum. In order to track the different integrability breaking phenomena reviewed sec.~\ref{sec:weak_chaos_koopman}, in this section, we will define known and new quantities in Krylov methods.

\paragraph{Krylov spread.}

Given an observable $g_0$, Lanczos applied to the Koopman generator $\mathcal K$ produces a Krylov basis $\{|n\rangle\}_{n\ge0}$ and a corresponding Krylov wavefunction $\phi_n(\sigma)$ with probabilities $p_n(\sigma)=|\phi_n(\sigma)|^2$. The Krylov spread rate,
\begin{equation}
C_K(\sigma) = \sum_{n\ge0} n\, p_n(\sigma)\,,
\end{equation}
measures the mean Krylov index explored at parameter $\sigma$. It quantifies the drift of spectral weight towards higher Krylov levels and provides a coarse measure of operator growth.

In weakly non-integrable Hamiltonian systems one should not expect coherent or ballistic spreading in the observable Hilbert space. Unlike quantum chaotic many-body systems, where Krylov growth often exhibits universal exponential or linear regimes, weak integrability breaking does not generically produce parametrically enhanced growth rates relative to the integrable case. Because the underlying phase space remains largely organised by surviving invariant tori, the deformation primarily induces slow redistribution of spectral weight through resonance-driven channels rather than uniform instability. As a result, the transport properties of the deformed system can differ qualitatively from the integrable limit but remain strongly observable dependent and non-universal. The Krylov evolution therefore reflects the local structure of phase space such as resonant layers and weakly chaotic bands rather than a global growth law, and provides a probe of how subtle integrability breaking reorganises dynamical directions in observable space.

\paragraph{Inverse participation ratio and delocalisation.}

For a chosen observable $g_0$, a complementary probe is provided by the inverse participation ratio,
\begin{equation}
\mathrm{IPR}^{g_0}(\sigma) = \sum_{n\ge0} p_n(\sigma)^2\,,
\end{equation}
which measures how concentrated the Krylov wavefunction remains. Its inverse, $N_{\rm eff}^{g_0}(\sigma)=1/\mathrm{IPR}^{g_0}(\sigma)$, can be interpreted as the effective number of Krylov basis states populated at parameter $\sigma$.

As reviewed in section \ref{sec:packetisation}, in integrable systems, the spectral measure $\mu_{g_0}$ is pure point, and the corresponding Krylov wavefunction typically remains supported on a relatively small subset of modes, even when $C_K(\sigma)$ grows. Weak breaking of integrability leads instead to a pronounced delocalisation in Krylov space: spectral weight spreads across many Lanczos directions, resulting in a systematic decay of the IPR. The IPR therefore isolates the transition from structured, quasi-periodic growth to genuinely delocalised operator dynamics.  

\paragraph{Spectral transport and Wasserstein distance.}

To characterise integrability breaking at the level of the Koopman spectrum itself, we compare spectral measures extracted from the Krylov tridiagonal matrix. For a given observable $g_0$, Lanczos yields a finite-dimensional tridiagonal matrix $T_N$ whose eigendecomposition
\begin{equation}
T_N = U\, \mathrm{diag}(\lambda_j)\, U^\dagger
\end{equation}
defines the discrete spectral measure
\begin{equation}\label{eq:Krylov_approx_spectralm}
\mu^{(N)}_{g_0}(\omega) = \sum_j w_j\, \delta(\omega-\lambda_j)\,,\quad w_j=|U_{0j}|^2\,.
\end{equation}
This measure encodes how $g_0$ decomposes into effective Koopman frequencies at resolution $N$ and provides a finite dimensional approximation of the spectral measure capturing the autocorrelator \eqref{eq:autocorrelator}.

Rather than relying solely on low-order moments of $\mu_{g_0}^{(N)}$, we quantify changes in spectral structure using the Wasserstein-1 distance. Given two measures $\mu_A$ and $\mu_B$ with cumulative distributions $F_A$ and $F_B$, the 1-Wasserstein distance is
\begin{equation}
W_1(\mu_A,\mu_B)=\int_{\mathbb{R}}|F_A(\omega)-F_B(\omega)|\,\mathrm d\omega\,.
\end{equation}
Unlike the spectral variance, $W_1$ is sensitive to how spectral weight is transported across frequencies, not merely to the overall width of the distribution. In this work we compute $W_1$ between the integrable and deformed spectral measures associated with the same observable,
\begin{equation}\label{eq:wasserstein_int_def}
W_1\!\left(\mu^{(N)}_{g_0,\mathrm{int}},\mu^{(N)}_{g_0,\mathrm{def}}\right).
\end{equation}

In addition to this raw transport, we also consider a shape transport in which trivial rigid shifts and overall rescalings of the spectrum are removed. Concretely, for a discrete measure $\mu^{(N)}(\omega)=\sum_j w_j\,\delta(\omega-\lambda_j)$ we define its weighted mean and variance
\begin{equation}
\langle \lambda\rangle_{\mu}=\sum_j w_j\lambda_j,\qquad 
\mathrm{Var}_{\mu}(\lambda)=\sum_j w_j\bigl(\lambda_j-\langle\lambda\rangle_{\mu}\bigr)^2,
\end{equation}
and introduce the normalised eigenvalues
\begin{equation}
\hat\lambda_j=\frac{\lambda_j-\langle\lambda\rangle_{\mu}}{\sqrt{\mathrm{Var}_{\mu}(\lambda)}}\,,
\end{equation}
which induce the normalised spectral measure
\begin{equation}
\hat\mu^{(N)}(\hat\omega)=\sum_j w_j\,\delta(\hat\omega-\hat\lambda_j)\,.
\end{equation}
We then compute the corresponding normalised Wasserstein distance
\begin{equation}\label{eq:W1normalised}
W_1\left(\hat\mu^{(N)}_{g_0,\mathrm{int}},\hat\mu^{(N)}_{g_0,\mathrm{def}}\right)\,,
\end{equation}
which isolates deformation-induced redistribution of spectral weight (shape change) from an overall drift of the spectrum.
A non-zero Wasserstein distance therefore directly reflects deformation-induced spectral transport: the minimal amount of spectral weight that must be moved, and the distance over which it must be transported, to deform the integrable spectrum into the non-integrable one.

As discussed in section \ref{sec:weak_chaos_koopman}, in near-integrable systems, spectral peaks associated with quasi-periodic motion are not immediately destroyed; instead, weight is gradually redistributed into nearby modes. Such redistribution can leave low-order moments nearly unchanged while producing a sizeable Wasserstein distance. For this reason, $W_1$ provides a probe of weak integrability breaking at the spectral level.

\paragraph{Sector-resolved Krylov leakage.}

The diagnostics discussed above probe how an observable spreads within its Krylov chain. In weakly non-integrable systems it is equally important to ask whether this spreading remains confined to observable directions already present in the integrable dynamics, or whether it drives transport into genuinely new directions. To isolate this effect we introduce a sector-resolved leakage diagnostic. To do so, we assume the system allows for a fixed decomposition of the observable space, as defined in 
\begin{equation}\label{eq:decomp_leakage2}
\mathcal H_\mathrm{obs} = \mathcal H_\mathrm{prot} \oplus \mathcal H_\mathrm{rest}\,,
\end{equation}
where $\mathcal H_{\rm prot}$ denotes a protected subspace associated with the integrable dynamics (e.g. the subspace spanned by observables depending only on the integrable degrees of freedom).
Given the decomposition \eqref{eq:decomp_leakage2}, let 
\begin{equation}
P_{\rm prot}:\mathcal H_{\rm obs}\to\mathcal H_{\rm prot}
\end{equation}
denote the orthogonal projector onto the protected sector. Projecting the generator to each sector yields two Krylov approximations of the spectral measures $\mu^{(\mathrm{prot})}$ and $\mu^{(\mathrm{rest})}$ as defined in eq.~\eqref{eq:Krylov_approx_spectralm}, and where we suppress the observable dependence not to overload the notation. To quantify the effective number of Koopman frequencies contributing within each sector, define the associated participation ratio is defined as
\begin{equation}\label{eq:PR}
\mathrm{PR}^{(\rm A)}_{\mu} =\frac{1}{\sum_k w_k^2}\,, \qquad A\in \{\mathrm{prot}\,,\mathrm{rest}\}\,,
\end{equation}
where the weights $w_j$ are defined in \eqref{eq:Krylov_approx_spectralm}.
To obtain a dimensionless measure of sector dominance we define the normalised fractions
\begin{equation}\label{eq:def_sector_fraction}
f_{\mathrm{prot}} = \frac{\mathrm{PR}_\mu^{(\mathrm{prot})}} {\mathrm{PR}_\mu^{(\mathrm{prot})} + \mathrm{PR}_\mu^{(\mathrm{rest})}}\,, \qquad f_{\mathrm{rest}} = 1-f_{\mathrm{prot}}\,.
\end{equation}
By construction, $f_{\mathrm{prot}}\approx 1$ indicates that spectral weight is concentrated in integrable directions, whereas a decrease of $f_{\mathrm{prot}}$ signals redistribution of spectral support into deformation-activated sectors. This sector-resolved spectral dominance complements the Krylov spread and Wasserstein diagnostics. Krylov delocalisation can occur predominantly within $\mathcal H_\mathrm{prot}$ without substantial cross-sector mixing, while a reduction of $f_{\mathrm{prot}}$ directly measures the relative growth of non-integrable channels.

A closely related manifestation of stretching and folding in classical string dynamics has been analysed explicitly in the context of chaotic and turbulent string evolution, where spatial structure is transferred to higher modes along the string worldsheet \cite{Ishii:2016rlk}. In that setting, filamentation is typically characterised through Fourier decomposition along the spatial direction, which reveals exponential growth of higher harmonics and power-law cascades in conserved quantities. In the present work, rather than resolving the dynamics in a fixed Fourier basis, we will characterise the redistribution of functional structure using a Krylov decomposition generated by the Koopman operator.

Note that $f_{\mathrm{prot}}$ is conceptually related to the notion of Krylov coherence (koherence) introduced in \cite{Balasubramanian:2025xkj}, which measures the entropy of overlaps between two distinct Krylov constructions and thus quantifies sensitivity to perturbations of the initial state or Hamiltonian. Rather than comparing nearby Krylov bases, here we track how spectral weight redistributes between protected and deformation-activated sectors.

\vspace{10pt}
Taken together, this choice of three diagnostics has as aim to  probe distinct and complementary aspects of near-integrable dynamics described in section \ref{sec:weak_chaos_koopman}. The Krylov spread rate $C_K$ quantifies how rapidly operators explore Krylov space. The IPR diagnoses whether this exploration remains structured or becomes delocalised. The Wasserstein distance measures how the underlying observable spectrum is reorganised by the deformation.

\subsection{The gEDMD approximation and numerical implementation} \label{sec:gEDMD}

In this section we briefly review the generator Extended Dynamic Mode Decomposition (gEDMD) approximation \cite{klus2020data,Williams2015} used throughout this work, and readers already familiar with gEDMD may skip this subsection without loss of continuity.  

\begin{figure}[t]
    \centering
    \includegraphics[width=1.0\linewidth]{ 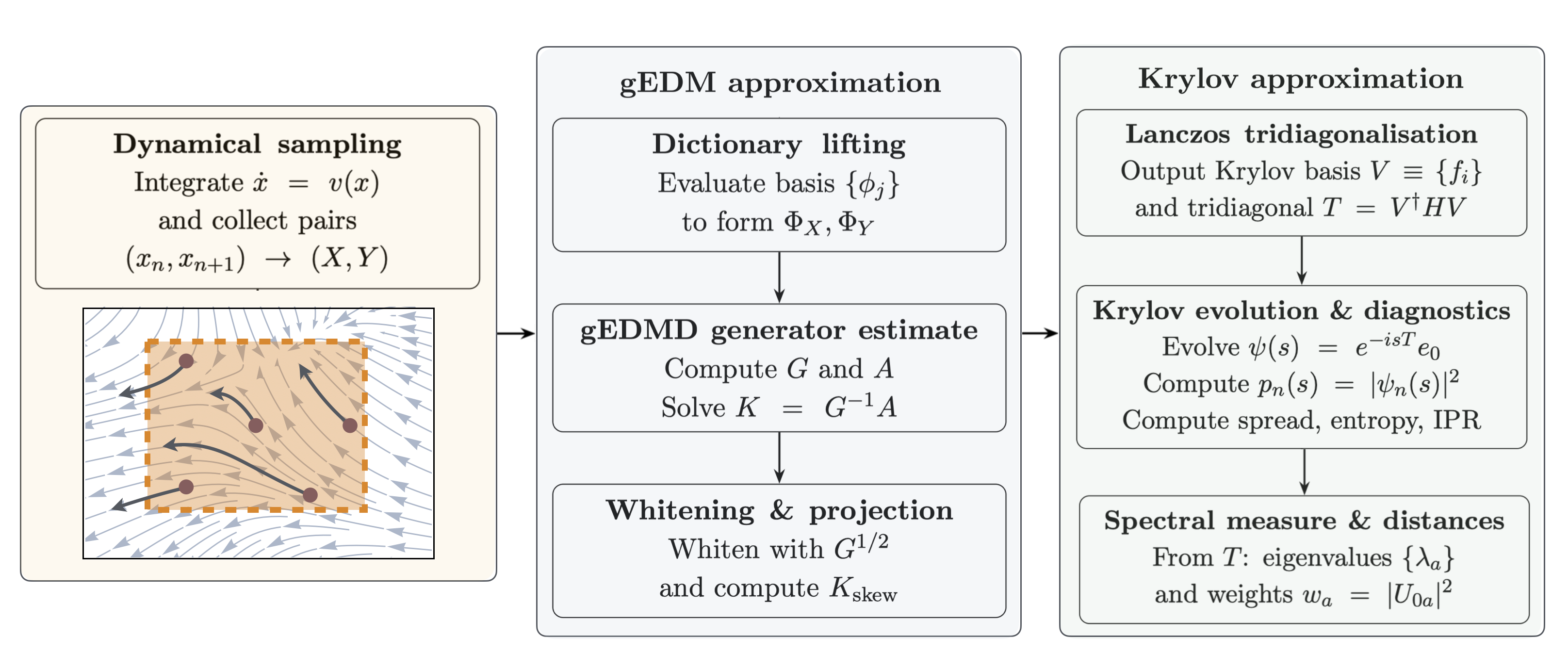}
    \caption{Schematic overview of the numerical pipeline: dynamical sampling of trajectories, gEDMD approximation of the Koopman generator in a finite dictionary, and Krylov-based evolution used to extract spectral diagnostics.}
    \label{fig:gEDMD_pipeline}
\end{figure}

As seen in sec.~\ref{sec:string_reductions}, the Koopman generator $\mathcal{L}=F\nabla$ provides a linear, infinite-dimensional representation of a generally nonlinear flow $\dot x = F(x)$ on phase space $X$.  Generator EDMD, i.e. gEDMD, produces a finite-dimensional approximation of $\mathcal{L}$ directly from the picked sample of trajectories: one specifies a finite basis of phase-space functions that we call ``dictionary'', samples the vector field along a collection of trajectories, and fits the best Galerkin-projected generator in that observable subspace. This is the object we use throughout as an effective finite-dimensional Liouvillian for Krylov constructions, as summarised schematically in fig.~\ref{fig:gEDMD_pipeline}.
For additional background and details on gEDMD and Koopman generators, see e.g. \cite{brunton2021modernkoopmantheorydynamical,korda2018linear,klus2020data}.

\paragraph{Observable space and reference measure.}
To define inner products and projections we equip the space of observables with a measure.  In the Koopman formulation of section~\ref{sec:koopman} the observable Hilbert space is $L^2(X,\mu)$, where $\mu$ is an invariant measure. In practice $\mu$ is not known analytically. Instead, given a finite cloud of trajectory data, we define the approximate inner-product
\begin{equation}\label{eq:measure_rho}
\langle f,g\rangle_\rho = \int_X f(x)^* g(x)\, \mathrm d\rho(x)\,.
\end{equation}
using an approximate measure from the finite data
\begin{equation}
    d\rho(x) = \frac{1}{M}\sum_{m=1}^M \delta(x-x_m)\,,
\end{equation}
and approximate the inner product determined by the measure $\mu$.

For Hamiltonian flow one has $\mathcal L = \{\cdot,H\}$ and $K=-i\mathcal L$. The Koopman generator $\mathcal{L}=F\nabla$ acts linearly on the space of observables $H_\mathrm{obs}$ but is infinite-dimensional and unbounded. To obtain a finite approximation, we choose a set of linearly independent observables
\begin{equation}
	\Phi(x)=\big(\phi_1(x),\ldots,\phi_N(x)\big)^{T}\,,\qquad
	\mathcal{D}_N=\mathrm{span}\{\phi_1,\ldots,\phi_N\}\subset \mathcal H_\mathrm{obs}\,,
\end{equation}
referred to as the dictionary.

Since $\mathcal{L}$ acts by differentiation along the flow, the dictionary should be approximately closed under this action. For the systems studied here, which are trigonometric in angular variables and polynomial in momenta, we use combinations of
\begin{align}
	&\{\sin(kq),\cos(kq)\}_{k=1}^K,\\
	&\{u^\alpha w^\beta z^\gamma : 1\le\alpha+\beta+\gamma\le p\},\\
	&\text{low-order mixed terms}\,.
\end{align}
We will comment later on the implications of a given choice and the fixed size of the dictionary.
Note that the constant observable is excluded.  Since $\mathcal{L}1=0$, it corresponds to a trivial zero mode and introduces a dominant static component in the Gram matrix \eqref{eq:gram_matrix_eval}. Equivalently, one may centre the dictionary by defining
\begin{equation}
	m=\frac{1}{M}\sum_{m=1}^M \Phi(x_m)\,,\qquad \tilde\Phi(x)=\Phi(x)-m\,,
\end{equation}
which restricts the approximation to fluctuations orthogonal to constants. 

\paragraph{Galerkin projection of the Koopman generator.}  
We approximate the action of $\mathcal{L}$ by its $\rho$-orthogonal projection onto the dictionary $\mathcal{D}_N$. 
Writing $\Pi_D\equiv \Pi_{\mathcal{D}_N}$ for this projection, we require
\begin{equation}
	\Pi_{N}(\mathcal{L}\phi_j)=\sum_{i=1}^N K_{N,ij}\phi_i\,,
\end{equation}
which defines a finite matrix $K_{N}$ representing the Liouville operator $\mathcal{L}$ on $\mathcal{D}_N$. The coefficients $K_{ij}$ are fixed by the Galerkin condition: for all $\phi_i\in\mathcal{D}_N$,
\begin{equation}
	\langle \phi_i,\mathcal{L}\phi_j\rangle_\rho=\sum_{k=1}^N K_{N,kj}\langle \phi_i,\phi_k\rangle_\rho\,,
\end{equation}
where the inner-product is given in \eqref{eq:measure_rho}.
Introducing the Gram matrix and action matrix
\begin{equation}\label{eq:gram_matrix}
	G_{ij}=\langle \phi_i,\phi_j\rangle_\rho\,,\qquad  A_{ij}=\langle \phi_i,\mathcal{L}\phi_j\rangle_\rho\,,
\end{equation}
the projected generator is $K=G^{-1}A$.
Thus, $K$ is the representation of $\mathcal{L}$ within the chosen dictionary.

\paragraph{Data-driven approximation from a cloud of trajectories.}
As mentioned earlier, in practice the measure $\rho$ in \eqref{eq:measure_rho} is not known explicitly.  Instead we approximate inner products using a set of samples $\{x_m\}_{m=1}^M$ obtained from the flow.
Concretely, we draw a set or cloud of initial conditions centred around a prescribed region of phase space, evolve each initial condition under the equations of motion, and collect a large number of snapshot pairs
\begin{equation}
	(x_m, x_m^+)\,,\qquad x_m^+=\varphi_{\Delta \sigma}(x_m)\,,
\end{equation}
obtained from numerically integrating the $\sigma$-flow $\dot x = F(x)$.(e.g. through Runge-Kutta methods).  These snapshot pairs provide empirical access to both the dictionary values $\phi_i(x_m)$ and the directional derivatives $\nabla\phi_j(x_m)F(x_m)$ along the vector field. 
We approximate the inner products in \eqref{eq:measure_rho} by an averaging over a cloud of pairs 
\begin{equation}
	\langle f\rangle_\rho \approx \frac{1}{M}\sum_{m=1}^M f(x_m)\,,
\end{equation}
yielding
\begin{equation}\label{eq:gram_matrix_eval}
	\hat G_{ij}=\frac{1}{M}\sum_{m=1}^M \phi_i(x_m)\phi_j(x_m)\,, \quad
	\hat A_{ij}=\frac{1}{M}\sum_{m=1}^M \phi_i(x_m)\,\big(\nabla\phi_j(x_m)F(x_m)\big)\,.
\end{equation}
To stabilise the inversion of the Gramm matrix $\hat G$, we introduce a small Tikhonov regularisation parameter $\lambda\ge0$ and define
\begin{equation}\label{eq:generator-EDMD_K}
	\hat K=(\hat G+\lambda I)^{-1}\hat A\,.
\end{equation}
This matrix is the generator-EDMD  approximation of $\mathcal{L}$.

\paragraph{Observable evolution and autocorrelator.}
Any observable $g\in \mathcal H_\mathrm{obs}$ can be written as $g(\sigma)=c^{T}\Phi(\sigma)$. 
Under Koopman evolution, its coefficient vector evolves approximately as
\begin{equation}
	\dot a(\sigma)=\hat K^{T}a(\sigma),\qquad a(0)=c\,,
\end{equation}
with $\hat K$ as define in \eqref{eq:generator-EDMD_K}.
Autocorrelation functions \eqref{eq:autocorrelator} are then approximated by
\begin{equation}
	C(\sigma)=\langle g,e^{\sigma\mathcal{L}}g\rangle_\rho \approx c^{T}\hat G\,e^{\sigma\hat K}c\,.
\end{equation}
In this work, $\hat K$ is used as an effective finite-dimensional Liouvillian (through a representation in the dictionary subspace) driving Krylov constructions, rather than as an eigenvalue solver: the diagnostics we employ depend on the induced finite-resolution spectral measures and on Krylov delocalisation, not on pointwise spectral convergence of $\hat K$.

\paragraph{Remarks.}

\begin{itemize}
    \item {\bf Finiteness of dictionary.} Our Koopman-Krylov diagnostics depend on two important choices that should be kept in mind when interpreting the results. In particular, in the numerical analysis carried in sec.~\ref{sec:KK_obs} we always make sure that 1) all quantities extracted from a given finite dictionary space are robust under changes of basis within that space, 2)  modifying the dictionary by small changes in choice of the functions spanning the dictionary space leave the diagnostic mostly invariant. 
    \item {\bf Introducing an energy-shell.}
Often, classical string dynamics exhibits a pronounced dependence of chaotic indicators on energy. This is seen as well in semiclassical non-integrable string dynamics and their Poincar\'e sections and Lyapunov exponents, see e.g. \cite{Panigrahi:2016zny,Basu:2011fw,PandoZayas:2010xpn}. This reflects the non-uniform nature of integrability breaking: at low energies motion may remain close to integrable, while at higher energies resonances proliferate and chaotic regions expand. 

The same consideration applies to our observable-based framework. Koopman spectra depend on the frequency content and resonance structure sampled by the ensemble of trajectories, which vary significantly across energy sectors. Mixing trajectories from different energies would blend qualitatively distinct spectral contributions and mask deformation-induced effects. To isolate these effects we restrict attention to ensembles drawn from a narrow energy shell of the reduced Hamiltonian,
\begin{equation}\label{eq:energy_window}
\mathcal{E}_{E_0,\Delta E} = \left\{x\in\Gamma \;\middle|\; E_0-\Delta E \le H(x) \le E_0+\Delta E \right\}\,,
\end{equation}
and compare integrable and deformed dynamics within this fixed regime. Since energy is conserved in the reduced system, this restriction ensures that the Koopman spectral measures being compared probe the same class of invariant or weakly broken structures in phase space. In this sense fixing an energy shell plays a role analogous to fixing the energy in Poincar\'e sections or Lyapunov exponent analyses, but adapted to our spectral, observable-based diagnostics.
    \item {\bf Numerical truncation.} We represent the evolved observable in the Krylov basis as a normalised wavefunction $\phi(\sigma)=\sum_{n=0}^{m-1}\phi_n(\sigma)$ such that $\sum_{n=0}^{m-1}|\phi_n(\sigma)|^2=1$
where $m$ is the Krylov-chain truncation dimension. Since the true dynamics lives in an infinite (or much larger) Krylov space, the truncated evolution is reliable only as long as the wavefunction has negligible support near the end of the chain. To quantify this, we define the edge mass of the Krylov wavefunction
\begin{equation}
\mathcal E_L(\sigma)=\sum_{n=m-L}^{m-1}|\psi_n(\sigma)|^2\,,
\end{equation}
i.e. the total probability contained in the last $L$ Krylov sites. For a fixed small $\epsilon$, we then define
\begin{equation}
\sigma_\star=\sup\{\,\sigma:\ \mathcal E_L(\sigma)<\varepsilon\,\}\,,
\end{equation}
and restrict all diagnostics to $\sigma\le \sigma_\star$. When $\mathcal E_L(\sigma)$ becomes non-negligible, finite-$m$ boundary effects can contaminate the apparent dynamics.
\end{itemize}

\section{Koopman-Krylov probes in integrability-breaking string solutions}\label{sec:KK_obs}

In this section we apply the Koopman-Krylov tools introduced in the previous section to three distinct non-integrable or weakly chaotic semiclassical string solutions. Rather than serving as a binary test for chaos, Koopman-Krylov diagnostics provide observable-resolved information on how phase-space dynamics reorganises under weakly non-integrable $\sigma$-flow. Our goal is to identify which features of observable spreading are universal across near-integrable string backgrounds, and which depend on the specific mechanism by which integrability is broken.

Before turning to specific backgrounds, it is useful to clarify how the results in this section will be interpreted. Throughout, we organise our analysis along three complementary axes. First, we assess observable sensitivity: whether a given observable couples directly to the structures introduced by the deformation, such as resonant terms or conserved combinations, or instead probes directions largely orthogonal to them. Second, we characterise the transport mechanism responsible for observable spreading, distinguishing between transport driven by resonance overlap, by the activation of additional dynamical degrees of freedom, or by an integrable enlargement of phase space that preserves invariant tori. Third, we examine the resulting spectral signature in Koopman-Krylov space, identifying whether the deformation induces a smooth spectral drift, a non-monotonic reshuffling of spectral weight, or near-rigidity of the effective generator.

\subsection{Landau-Lifshitz limit of the two-loop  \texorpdfstring{$SU(2)$}{SU(2)}-dilation operator}\label{sec:su2_2loop}
The dilatation operator corresponding to the $SU(2)$ sector of planar $\mathcal N=4$ SYM provides a natural setting in which non-integrable dynamics emerge within an otherwise integrable framework. At one loop, it coincides with the Heisenberg $SU(2)$ spin chain and is thus integrable \cite{Minahan:2002ve}. Particularly interesting is its continuum description in terms of the Landau-Lifshitz (LL) sigma model.  The continuum Landau-Lifshitz description arises from the long-wavelength limit of the $SU(2)$ spin chain and admits a direct correspondence with the action describing semiclassical string dynamics on $\mathrm{AdS}_5\times S^5$ \cite{Kruczenski:2004cn,Kruczenski:2004kw,Stepanchuk:2012xi}. In particular, one considers strings moving on an $S^3\subset S^5$ subspace and localised at the centre of $\mathrm{AdS}_5$. In this regime the classical spin-chain equations of motion coincide with those of the string sigma model after taking a suitable non-relativistic limit of the effective coupling \cite{Kruczenski:2003gt}. The classical Landau-Lifshitz model thus captures the long-wavelength, semiclassical limit of the quantum spin chain, providing a direct bridge between the gauge-theory dilatation operator and semiclassical string dynamics developed in \cite{Engquist:2003rn,Beisert:2003xu}.

More precisely, the coherent-state path-integral formulation, these higher-loop effects translate into higher-derivative and nonlinear corrections to the Landau-Lifshitz action, becomes a higher derivative action
\begin{equation}\label{eq:LLact_su2}
    S = L \int dt \int_{0}^{2\pi} \frac{\mathrm d\sigma}{2\pi} \left( \cos(2\psi) \dot{\phi} - H[\vec{n}] \right)\,,
\end{equation}
with Hamiltonian given by 
\begin{equation}
    H[\vec{n}] = a_0 (\partial_\sigma \vec{n})^2 + a_1 (\partial_\sigma^2 \vec{n})^2 + a_2 (\partial_\sigma \vec{n})^4\,,
\end{equation}
where the higher-derivative terms are then continuum equivalent of the  longer-range and multi-spin interactions of the chain interpretation of the dilation operator\footnote{As already remarked in the introduction, planar $\mathcal N=4$ SYM is strongly believed to admit an integrable all-loop spectral description. However, there is no known closed-form finite-length expression for the full all-loop dilatation operator. Instead, one works either with perturbative long-range Hamiltonians obtained order by order in the coupling, or with asymptotic Bethe methods. Truncations of the perturbative Hamiltonian generally break integrability, even though formal long-range integrable completions exist, such as the class constructed in \cite{Bargheer:2008jt} and one is confronted with the non-integrable theories.}. The coefficients $a_1$ and $a_2$ parametrise integrability-breaking deformations away from the tree-level Landau-Lifshitz action. These constants are related to the 't Hooft coupling $\lambda$ and the length of the spin chain $L$ via
\begin{equation}
    a_0 = \frac{1}{8}\,, \quad  a_1 = -\frac{\lambda}{32L^2}\,, \quad  a_2 = \frac{3\lambda}{128L^2}\,,
\end{equation}
where $\lambda_{\rm eff}\sim \lambda/L^2$ controls the strength of higher-loop corrections. 

This model is known \cite{McLoughlin:2022jyt} to exhibit weak chaotic behaviour: the integrable structure is perturbatively broken, and Lyapunov exponents become non-zero, though small. It therefore provides a controlled playground in which to test the sensitivity of Koopman-Krylov diagnostics. We will restrict the analysis to the same solution Ansatz as \cite{McLoughlin:2022jyt}, which is given by
\begin{equation} \label{eq:ansatzsu2}
    \phi(\sigma,t) = \omega t\,, \quad  \psi(\sigma,t) = \frac{1}{2} q(\sigma)\,.
\end{equation}
describing a circular strings and the system reduces to a one-dimensional mechanical model in terms of the variables $q$. The resulting Hamiltonian is written in Ostrogradsky form, treating $q$ and its first derivative $q^{(1)}$ as independent phase-space coordinates with conjugate momenta $p_0,p_1$,
\begin{equation}\label{eq:Ham_su2_ch}
    H_{\mathrm{eff}} = q^{(1)} p_0 - \frac{p_1^2}{4a_1} + a_0 \left(q^{(1)}\right)^2 + (a_1 + a_2) \left(q^{(1)}\right)^4 - \omega \cos q\,.
\end{equation}

\subsubsection{Koopman generator for the circular $SU(2)$ sector}\label{sec:koopmansu2}
We now derive the Koopman operators for the integrable one-loop Hamiltonian reduction and its two-loop deformation.
The equations of motion associated to the Hamiltonian \eqref{eq:Ham_su2_ch} are a fourth-order ODE in $\sigma$ 
\begin{equation}\label{eq:def_circul_string_eoms}
	a_0\,q^{(2)} + 6(a_1+a_2)\,(q^{(1)})^{2}\,q^{(2)}+a_1\,q^{(4)}-\omega \sin q = 0\,,
\end{equation}
where $\bullet{}^{(n)}\!=\partial_\sigma^n$. As before we introduce first- order state vector $x(\sigma)=(q,u,w,z)=\big(q,\;q^{(1)},\;q^{(2)},\;q^{(3)}\big).$ Then the differential equation is equivalent to the first-order system
\begin{gather}
	\begin{aligned}
q_\sigma = u\,,\quad u_\sigma = w\,,\quad w_\sigma = z\,,\quad  z_\sigma = \frac{a_0}{a_1}\,w +6\Big(1+\frac{a_2}{a_1}\Big)u^2 w +\frac{\omega}{a_1}\,\sin q\,.
\end{aligned}
\end{gather}
This defines a vector field $F(x)$ on the 4D state space, $x' = F(x)$, where 
\begin{equation}
    F = (u,\, w,\, z,\, a_1^{-1}(\omega \sin q - a_0 w - 6(a_1+a_2)u^2 w))\,.
\end{equation}
The Koopman generator acting on observables $g(x)$ is the Lie derivative along $F$:
\begin{equation}\label{eq:Koopman_su2_ch}
	\mathcal L_\sigma^{(\mathrm{def})} = u\,\partial_q + w\,\partial_u + z\,\partial_w +	\left(\frac{a_0}{a_1}\,w + 6\left(1+\frac{a_2}{a_1}\right)u^2 w - \frac{\omega}{a_1}\sin q\right)\partial_z\,.
\end{equation}
For any observable $g(x)$, its $\sigma$-autocorrelator \eqref{eq:autocorrelator}. Note that since $a_1=a_2=0$, the equation reduces to the pendulum form
\begin{equation}\label{eq:integrable_eoms}
	a_0\,q^{(2)}-\omega\sin q=0\,.
\end{equation}
Hence becoming a 2D first-order system given by
\begin{equation}
	q_\sigma=u\,,\qquad u_\sigma=\frac{\omega}{a_0}\sin q\, ,
\end{equation}
and the Koopman generator reads
\begin{equation}\label{eq:Koopman_su2_int}
	\mathcal L_\sigma^{(\mathrm{int})} = u\,\partial_q + \frac{\omega}{a_0}\sin q\,\partial_u\,.
\end{equation}

A crucial point to note is that for this particular case, the integrable truncation is effectively two-dimensional determined by 
\eqref{eq:Koopman_su2_int}, whereas the deformed (Ostrogradsky) dynamics in \eqref{eq:Koopman_su2_ch} lives on a four-dimensional phase space.  For the comparisons below, it is useful to view the integrable flow as embedded into the same four-dimensional space by restricting to an invariant submanifold (the additional coordinates and their conjugate momenta set to zero, consistently preserved by the undeformed equations).

\begin{figure}[t]
    \centering
    \includegraphics[width=0.7\linewidth]{ 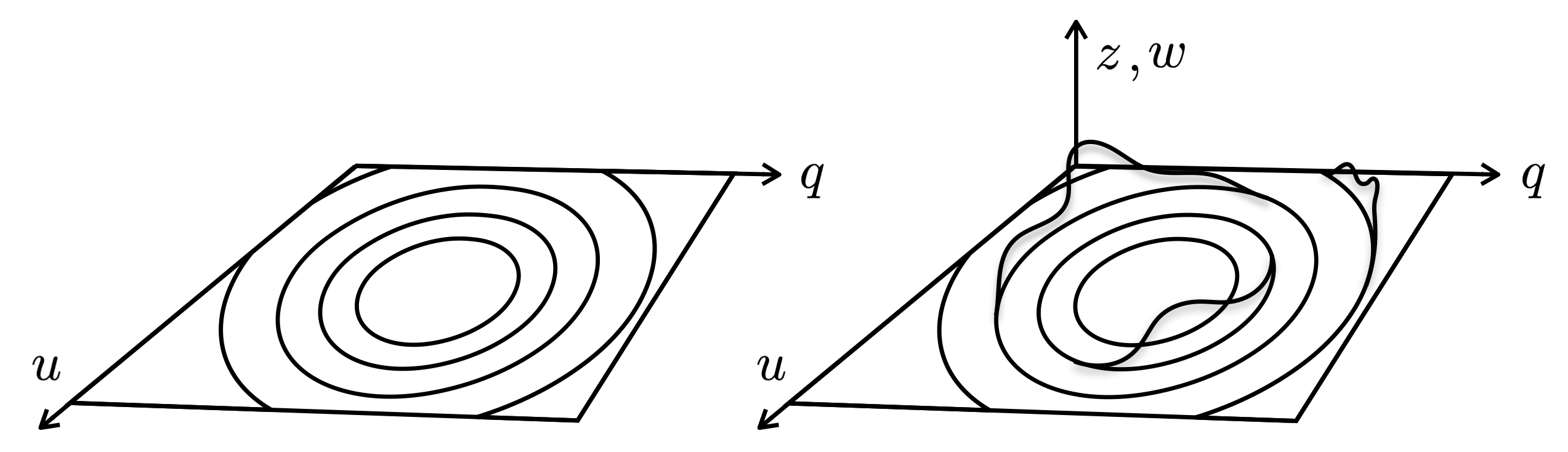}
    \caption{Schematic illustration of how weak integrability breaking introduces, in classes of system as the higher-derivative $su(2)$-sector, additional dynamical directions and, hence, resonant channels. Left: in the integrable truncation, motion is confined to invariant action-angle tori and observables decompose into discrete quasi-periodic frequencies.
Right: the deformation activates additional degrees of freedom, allowing slow, resonance-driven coupling between otherwise decoupled sectors.}
    \label{fig:breaking}
\end{figure}

\subsubsection{Choice of observables and sector decomposition}
Because of the nature of classical integrability breaking reviewed in sec.~\ref{sec:weak_chaos_koopman}, Koopman-Krylov diagnostics are probe dependent, the choice of observable is part of the definition of what is being compared. Moreover, in this particular case, the integrable and deformed truncations live on different phase spaces: the integrable system closes on the longitudinal variables $(q,u)$, whereas the deformation enlarges phase space by transverse degrees of freedom $(w,z)$. We therefore compare paired observables $(g^{(\mathrm{int})},g^{(\mathrm{def})})$ rather than identical functions.

Concretely, each pair is chosen so that $g^{(\mathrm{def})}$ reduces to $g^{(\mathrm{int})}$ upon restriction to the longitudinal submanifold $(w,z)=0$, while allowing additional terms in the deformed case that isolate or couple to the transverse sector. We take
\begin{align}
g_E^{(\mathrm{def})}(q,u,w,z) &\equiv E= \tfrac12 a_0 u^2 - \omega \cos q + \kappa w^2\,, 
&\quad g_E^{(\mathrm{int})}(q,u) &\equiv \tfrac12 a_0 u^2 - \omega \cos q\,,\nonumber\\
g_{E_\perp}^{(\mathrm{def})}(q,u,w,z) &\equiv E_\perp= \tfrac12 (w^2+z^2)\,,
&\quad g_{E_\perp}^{(\mathrm{int})}(q,u) &\equiv \tfrac12 a_0 u^2\,,\\
g_P^{(\mathrm{def})}(q,u,w,z) &\equiv P= u\cos q + w\sin q\,,
&\quad g_P^{(\mathrm{int})}(q,u) &\equiv u\cos q\,. \nonumber
\end{align}
Here $E$ is the conserved reduced energy (with the deformation-induced contribution explicit), $E_\perp$ isolates the transverse sector opened by the deformation, and $P$ probes longitudinal-transverse mixing. All Koopman-Krylov quantities shown below are computed separately for the integrable and deformed generators using their corresponding paired observables, and then plotted together to quantify deformation-induced spectral transport and sector leakage.

Note that these observables (and those defined in the other systems) are physically motivated but not canonical, and alternative choices would probe different aspects of the dynamics.

\subsubsection{Koopman-Krylov probes of weak classical chaos and resonance}
We now perform an observable-resolved comparison between the integrable and deformed dynamics in eqs.~\eqref{eq:Koopman_su2_ch} and~\eqref{eq:Koopman_su2_int}. For each observable $g$ introduced above, we construct separate gEDMD approximations of the Koopman generator for the integrable and deformed systems. This yields two autocorrelators
$C_g^{(\mathrm{int})}(\sigma)$ and $C_g^{(\mathrm{def})}(\sigma)$ and the associated spectral measures. From each pair we compute the same Krylov diagnostics defined in sec.~\ref{sec:Krylovtoolkit} and plot the integrable and deformed results together\footnote{Contrary to the systems discussed in the subsequent sections, in this system we do not impose the energy windowing procedure of eq.~\eqref{eq:energy_window}, as the deformation-induced spectral signatures are already clearly visible without further restriction of the sampling region.}. 

In the two-loop SU(2) Landau-Lifshitz system this comparison shows that integrability breaking manifests as resonance-driven spectral transport and observable-dependent leakage into higher-derivative degrees of freedom, producing Krylov delocalisation without global chaos. Appendix~\ref{sec:su2_action_angle_resonance} provides a perturbative KAM analysis supporting this interpretation.

\begin{figure}[t]
    \centering
    \includegraphics[width=1.0\linewidth]{   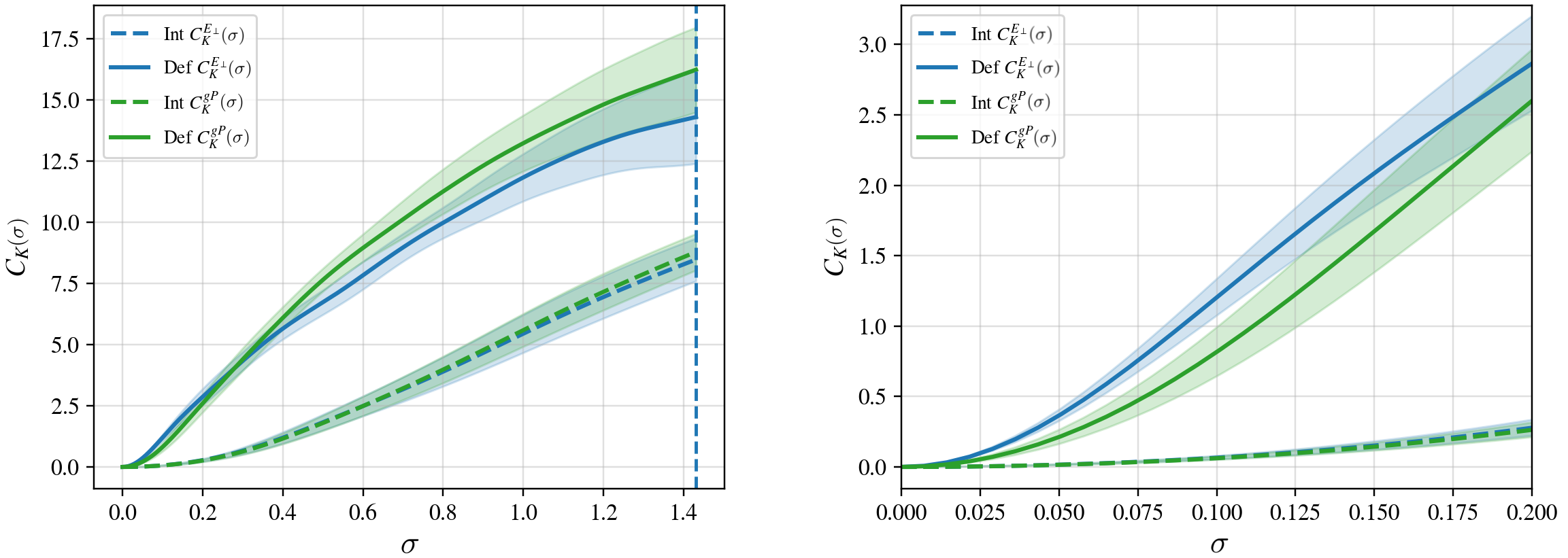}
    \caption{Left: Krylov complexity $C_K(\sigma)$ for two representative observables, the sector-resolved energy $g_E$ and the momentum-type probe $g_P$, comparing the integrable truncation (Int, in dashed lines) with the two-loop deformed system (Def, in solid lines) at fixed deformation point $(a_1,a_2)= (0.2,-0.6)$.
Right: small-$\sigma$ zoom highlighting the onset of delocalisation.
Shaded bands indicate the empirical uncertainty from the ensemble of initial conditions.}
\label{fig:su2_ck_gE_gP}
\end{figure}

\begin{figure}[t]
    \centering
    \includegraphics[width=1.0\linewidth]{   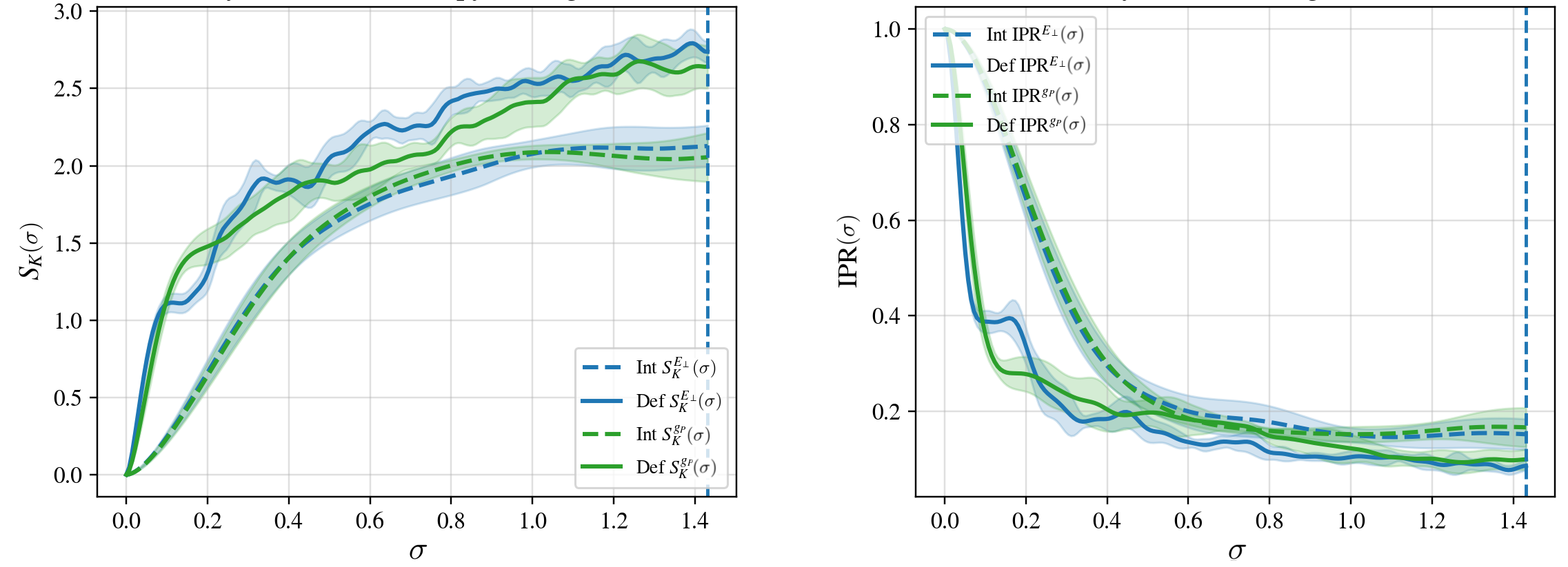}
    \caption{Left: Shannon entropy $S_K(\sigma)$, and right: Krylov inverse participation ratio $\mathrm{IPR}_K(\sigma)$ for the integrable truncation (Int, in dashed lines) and for the two-loop deformed $SU(2)$ Landau-Lifshitz model (Def, in solid lines), evaluated at representative deformation point $(a_1,a_2)=(0.2,-0.6)$.
We show two observables: the sector-resolved energy probe $g_{E_\perp}$ and the momentum-type probe $g_P$.}
\label{fig:su2_entropy_ipr}
\end{figure}

\paragraph{Spread and delocalisation in Krylov space.}
Turning first to the Krylov spread, figure~\ref{fig:su2_ck_gE_gP} shows that the two-loop deformation enhances Krylov growth relative to the integrable SU(2)-sector, both for energy-probes $g_{E_\perp}$ and for the momentum-type observable $g_P$.
Figure~\ref{fig:su2_entropy_ipr} further strengthens this interpretation by showing that the deformation increases the Krylov entropy $S_K(\sigma)$ and decreases $\mathrm{IPR}_K(\sigma)$. 
Thus, in this near-integrable regime, the deformation does not merely increase the mean Krylov depth, but also induces a genuine broadening of the Krylov distribution.
Importantly, the magnitude and onset of delocalisation remain strongly observable dependent, consistent with a picture in which integrability breaking enters through specific deformation-activated channels rather than through global chaotic mixing. Importantly, within the $\sigma$-range accessible to our numerics we do not observe a qualitative change in the growth law of the Krylov spread, nor any clear indication of early saturation. The deformation enhances delocalisation in a quantitative sense, but does not produce parametrically faster growth compared to the integrable case. This is consistent with the fact that we are probing the onset of weak integrability breaking rather than fully developed chaos. In contrast to quantum chaotic systems, where Krylov complexity can exhibit sharp universal growth regimes, the present semiclassical dynamics displays a comparatively mild and strongly observable-dependent redistribution in Krylov space.

\begin{figure}[t]
    \centering
    \includegraphics[width=0.95\linewidth]{   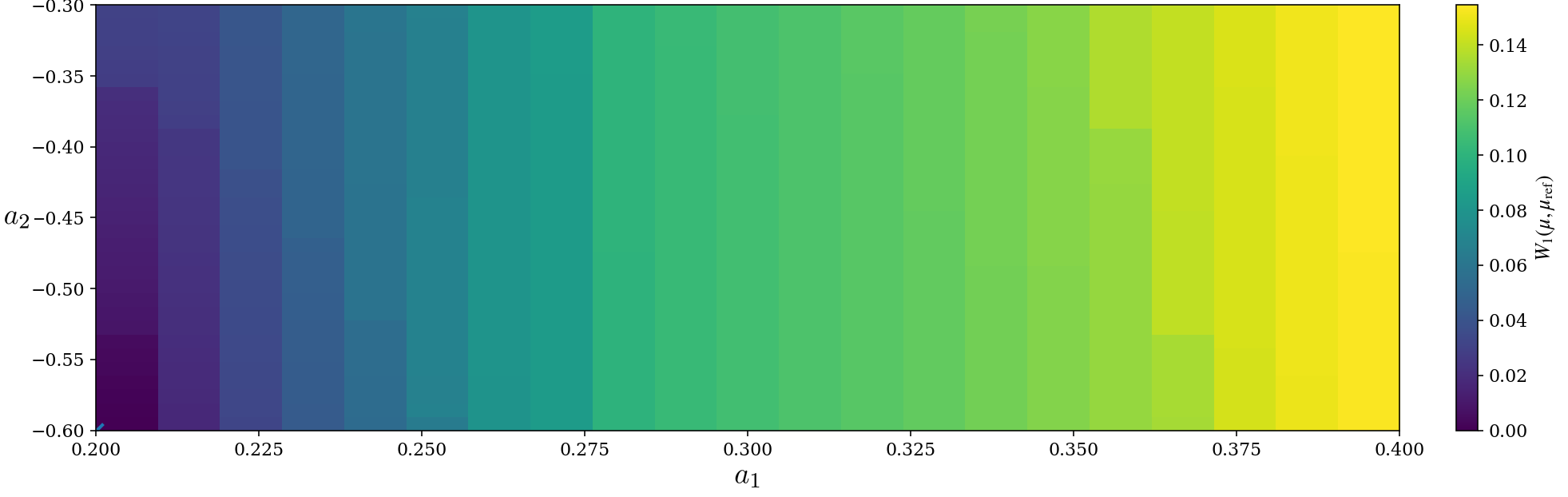}
    \caption{Heatmap of the median Wasserstein distance $W_1(\mu_{\rm int},\mu_{\rm def})$ between Koopman spectral measures in the integrable truncation and in the deformed theory, evaluated across a grid of deformation parameters $(a_1,a_2)$ and aggregated over the observable set $\{g_E,g_{E_\perp},g_P\}$ (median over observables at each point).}
\label{fig:su2_W1_heatmap_median}
\end{figure}

\paragraph{Spectral transport.}
To quantify how the deformation reshapes the Koopman spectrum in an observable-resolved manner, we compare spectral measures $\mu_g^{(\rm int)}$ and $\mu_g^{(\rm def)}$ using the 1-Wasserstein distance, as defined in eq.~\eqref{eq:wasserstein_int_def}.
This diagnostic is sensitive to  spectral transport: rather than requiring exponential sensitivity or global mixing, it measures how spectral weight is redistributed away from the quasi-discrete peak structure characteristic of integrable motion and into broader frequency components that arise from near-resonant forcing. Figure~\ref{fig:su2_W1_heatmap_median} summarises this effect across deformation space by plotting the median value of $W_1$ over the observable set $\{g_E,g_{E_\perp},g_P\}$.
Across the scanned range the distance grows systematically as $a_1$ is increased, indicating that the higher-derivative deformation term drives a robust increase in observable-resolved spectral redistribution.
The weaker variation along $a_2$ shows that, for this diagnostic and parameter range, quartic-gradient corrections have a subdominant effect compared to varying $a_1$. Appendix~\ref{sec:su2_action_angle_resonance} gives a perturbative explanation using KAM theory arguments and corroborates the trends seen in fig.~\ref{fig:su2_W1_heatmap_median}: varying $a_1$ shifts the linear frequencies across low-order resonances, whereas $a_2$ does not, consistent with the weaker $a_2$-dependence of $W_1$.

\paragraph{Protected-sector leakage.}
To isolate how much of the deformed dynamics remains supported on the integrable part of the phases space versus leaking into higher-derivative degrees of freedom, we use the sector-resolved dominance ratio $f_{\rm prot}$ defined above. This quantity is close to one when the spectral weight (as measured by the associated participation ratio $\mathrm{PR}_\mu$ defined in \eqref{eq:PR}) remains concentrated in the protected sector, and decreases as weight is transported into the phase space-part generated from the deformation.

\begin{figure}[t]
    \centering
    \includegraphics[width=1\linewidth]{   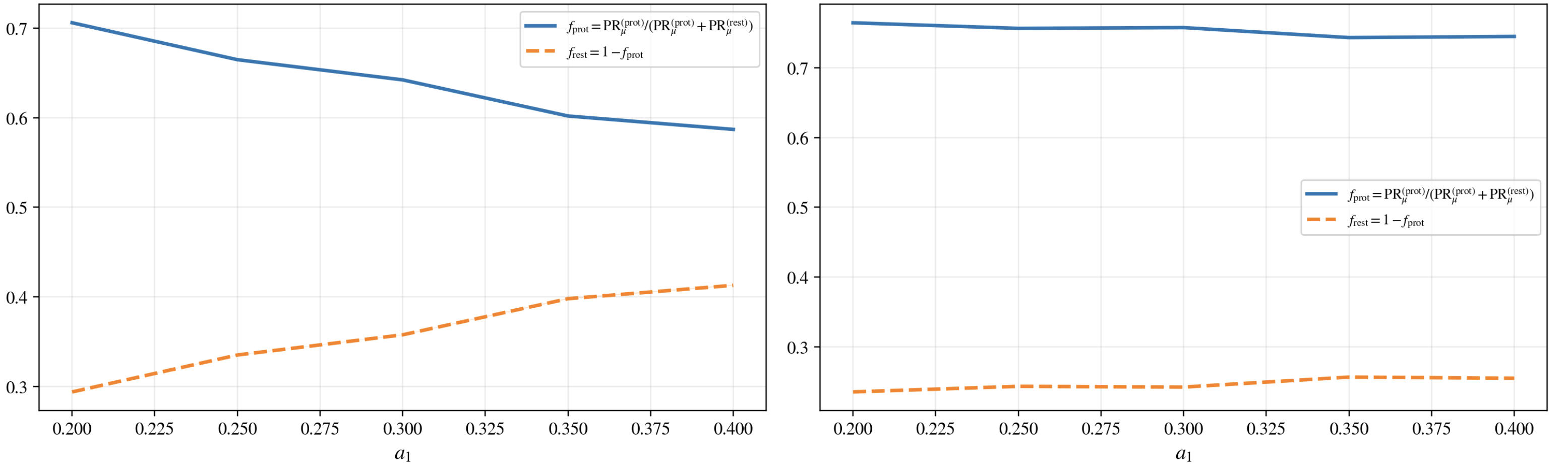}
    \caption{The normalised protected-sector dominance $f_{\rm prot}$ (solid) and the complementary leakage fraction $f_{\rm rest}=1-f_{\rm prot}$ (dashed) as functions of the deformation parameter $a_1$. The shaded band shows the variation over $a_2$ across the scan range (6-95\% quantiles). Left: $g_{E_\perp}$ exhibits increasing leakage into the non-protected sector as $a_1$ grows. Right: $g_P$ remains largely confined to the protected sector, demonstrating strong observable dependence of deformation-induced sector transport.}
\label{fig:su2_mprot_leakage_gEperp_gP}
\end{figure}

Figure~\ref{fig:su2_mprot_leakage_gEperp_gP} shows these ratio for the the observables $g_{E_\perp}$ and $g_P$. For the transverse-energy probe $g_{E_\perp}$, the protected-sector dominance decreases steadily as $a_1$ is increased: across the scanned range $f_{\rm prot}$ drops. This indicates that the higher-derivative deformation drives a weak, but systematic, leakage of the $g_{E_\perp}$-resolved dynamics into the non-protected sector, consistent with  formation of resonance-assisted excitation of the additional derivative modes rather than uniform chaotic mixing. By contrast, for the momentum-type observable $g_P$ the dominance ratio remains essentially flat ($f_{\rm prot}\approx0.75$ with only weak drift), showing that this probe largely stays confined to the protected sector even as $a_1$ is dialled.

Taken together, these results support the picture suggested by the spread or delocalisation  diagnostics discussed earlier: the deformation induces delocalisation in Krylov space while the underlying phase-space motion remains near-integrable, and the resulting ``filamentation'' into higher-derivative directions is strongly observable dependent.

\subsection{LL limit of the Leigh-Strassler-deformed  \texorpdfstring{$SU(3)$}{SU(3)}-dilatation operator} 
\label{sec:koopman_chaossu3}

In this section we study integrability breaking in the Leigh-Strassler deformation of planar $\mathcal N=4$ SYM. This deformation depends on two complex parameters and, for a specific choice of parameters, interpolates between an exactly integrable locus and a generically non-integrable regime. The general Leigh-Strassler deformation introduces two complex parameters $(h,q)$ through the superpotential
\begin{equation}
W = g''\, \mathrm{Tr} \left[ \Phi_0 \Phi_1 \Phi_2 - q \Phi_1 \Phi_0 \Phi_2 + \frac{h}{3}\left( \Phi_0^3 + \Phi_1^3 + \Phi_2^3 \right) \right]\,.
\end{equation}
As reviewed in \cite{Bundzik:2005zg}, the one-loop planar dilatation operator closes in the $SU(3)$ sector of chiral operators. It is convenient to parametrise
\begin{equation}
q = e^{\kappa + i\beta}, \qquad \kappa,\beta \in \mathbb R\,,
\end{equation}
where $\beta$ represents a phase twist and $\kappa$ a real anisotropy.  A distinguished integrable locus arises for $h=0\,, q=e^{i\beta}$ corresponding to the real $\beta$-deformation. On the string side this theory is dual to the Lunin-Maldacena background obtained by a TsT transformation (T-duality-shift-T-duality) of $\mathrm{AdS}_5\times S^5$ \cite{Lunin:2005jy,Frolov:2005dj}. The associated spin chain remains integrable, while turning on $\kappa\neq 0$ generically destroys integrability \cite{Berenstein:2004ys,Freyhult:2005ws,Beisert:2005if}. We restrict to $h=0$ and consider the weakly deformed regime $|\beta|,\;|\kappa|\ll 1$, so that the dynamics realises a small non-integrable perturbation of an integrable background.

The semiclassical dynamics of this sector is captured by the coherent-state Landau-Lifshitz action derived in \cite{Frolov:2005iq,Chen:2006bh}. We follow closely the derivation of \cite{McLoughlin:2022jyt}. In first-order form,
\begin{equation}
S = L \int \mathrm dt \int_0^{2\pi}\frac{\mathrm d\sigma}{2\pi} \Big( \cos^2\theta\,\dot\phi + \cos 2\psi\,\sin^2\theta\,\dot\varphi -  H[\vec n] \Big)\,,
\end{equation}
where $H[\vec n]$ is a nonlinear sigma-model Hamiltonian density depending on $(\theta,\psi,\phi,\varphi)$ and their spatial derivatives.

In the undeformed limit $\beta=\kappa=0$, the Hamiltonian density reduces to
\begin{equation}
H_0 = (\theta')^2 +\sin^2\theta\Big[ (\psi')^2 +\cos^2\theta\big(\phi'-\cos 2\psi\,\varphi'\big)^2 +\sin^2 2\psi\,(\varphi')^2 \Big]\,,
\end{equation}
which is integrable and admits a separation of variables analogous to the rigid $SU(2)$ sector.

To obtain a finite-dimensional Hamiltonian system we impose the rigidly rotating ansatz of \cite{McLoughlin:2022jyt},
which is compatible with the deformed equations of motion only for $\beta=0$,
\begin{equation}\label{Ansatzsu3}
\phi(\sigma,t)=\omega_\phi t\,,\quad \varphi(\sigma,t)=\omega_\varphi t\,,\quad \theta=\theta(\sigma)\,,\quad \psi=\psi(\sigma)\,,
\end{equation}
so that spatial derivatives of $\phi,\varphi$ vanish while their time derivatives remain fixed. After a Legendre transform with respect to $\sigma$, the undeformed reduced Hamiltonian density becomes
\begin{equation}\label{eq:su3_H0}
 H_{0,\mathrm{eff}} = p_\theta^2 + p_\psi^2\,\csc^2\theta - 4\omega_\phi \cos^2\theta - 4\omega_\varphi \cos 2\psi\,\sin^2\theta\,.
\end{equation}
This defines a two-degree-of-freedom integrable Hamiltonian system whose phase-space structure is organised by invariant tori.

The deformation introduces additional interaction terms. Most importantly, one obtains an explicit sextic potential \cite{Frolov:2005iq}
\begin{equation}
V_6(\theta,\psi) = \frac{9}{4} \left({\beta}^2+{\kappa}^2\right) \cos^2\theta\,\sin^4\theta\,\sin^2 2\psi\,,
\end{equation}
which has no analogue in the $SU(2)$ higher-derivative model. For $\beta=0$ this term couples the angular variables non-separably and provides the leading mechanism for $\kappa$-driven integrability breaking in the reduced dynamics. The total reduced Hamiltonian density therefore reads \cite{McLoughlin:2022jyt}
\begin{gather}
\begin{aligned}\label{eq:su3_Hdef}
H_{\kappa,\rm eff} &= p_\theta^2 + p_\psi^2\,\csc^2\theta - 4\omega_\phi \cos^2\theta - 4\omega_\varphi \cos 2\psi\,\sin^2\theta \\
&\quad + \frac{9}{4} \left({\beta}^2+{\kappa}^2\right) \cos^2\theta\,\sin^4\theta\,\sin^2 2\psi + \mathcal O(\beta\,\partial_\sigma, \kappa\,\partial_\sigma)\,.
\end{aligned}    
\end{gather}
For sufficiently small deformation parameters this system realises a weakly non-integrable perturbation of an integrable Hamiltonian with two degrees of freedom. In \cite{McLoughlin:2022jyt}, the authors showed that the semiclassical Landau-Lifshitz models arising from Leigh-Strassler deformations exhibit deformation-dependent chaotic behaviour through non-zero characteristic Lyapunov exponents.

Finally, the $\sigma$-flow generated by $H_{\kappa,\mathrm{eff}}$ defines a Koopman operator. As reviewed in section~\ref{sec:koopman}, the associated Koopman generator along the flow, through eq. \eqref{eq:Koopmanop}, which we represent in a finite dictionary for the Koopman-Krylov analysis.

\subsubsection{Choice of observables}
In contrast, the deformation discussed in sec. \ref{sec:su2_2loop}, the continuum limit of the integrable $SU(3)$-sector and its Leigh-Strassler deformation define two Hamiltonians on the same reduced phase space with coordinates $(\theta,\psi,p_\theta,p_\psi)$. We therefore use a single set of observables $g$ throughout, and compare the Koopman-Krylov diagnostics obtained by evolving the same $g$ with either the deformed or undeformed dynamics. Contrary to the previous example, all computations here are performed on a fixed energy window around a fix Hamiltonian energy around a value $E_0$, taken to be the same for the integrable and deformed Hamiltonians. In fact we will see that the spectral transport has a pronounced energy-dependence.

We take the following set of observables:
\begin{gather}
\begin{aligned}\label{eq:su3_observables}
g_V(\theta,\psi)&\equiv\sin^4\theta,\sin^2\psi,\cos^2\psi+\sin^2\theta,\cos^2\theta\,,\\
g_6(\theta,\psi)&\equiv
\cos^2\theta\,\sin^4\theta\,\sin^2\psi\,\cos^2\psi\,,\\
g_{\rm mix}(\theta,\psi,p_\theta,p_\psi)&\equiv-2p_\theta\sin(2\theta)\cos(2\psi)+p_\psi\sin(2\psi)\bigl(3\cos(2\theta)+1\bigr)\,,\\
g_K(\theta,\psi,p_\theta,p_\psi)&\equiv\frac12\Bigl(p_\theta^2 + p_\psi^2\csc^2\theta\Bigr)
\,.
\end{aligned}
\end{gather}
Here $g_V$ probes the angular potential structure independently of the overall deformation scale, while $g_6$ isolates the specific sextic angular profile multiplying the higher-order Hamiltonian term. The observable $g_{\rm mix}$ mirrors the explicit momentum-angle couplings introduced by the deformation and is therefore maximally sensitive to resonant channel activation. Finally, $g_K$ probes the kinetic part of the dynamics directly and tracks redistribution of spectral weight into momentum directions. All Koopman-Krylov quantities shown below are computed separately for the integrable and deformed generators using the same seed observables, and then compared to quantify deformation-induced spectral transport and sector reorganisation. As in the previous system, these probes are physically motivated but not canonical; alternative choices would emphasise different dynamical channels.

\subsubsection{Koopman-Krylov probes of weak classical chaos and resonance}
In this case, the integrable and deformed Hamiltonians dynamics live on the same reduced phase space, and integrability breaking reorganises trajectories primarily through resonant couplings rather than through an enlargement of accessible directions. As a result, Koopman-Krylov signatures are more subdued and strongly energy dependent. Qualitative differences in Krylov spread and delocalisation become visible only after restricting to appropriate energy windows that sample resonant regions of phase space. This observation mirrors the strong initial condition dependence of the Lyapunov characteristic exponents computed for this model in \cite{McLoughlin:2022jyt}. Within such windows we observe systematic but mild spectral transport and observable-dependent delocalisation. 

\begin{figure}[t]
    \centering
    \includegraphics[width=1.0\linewidth]{    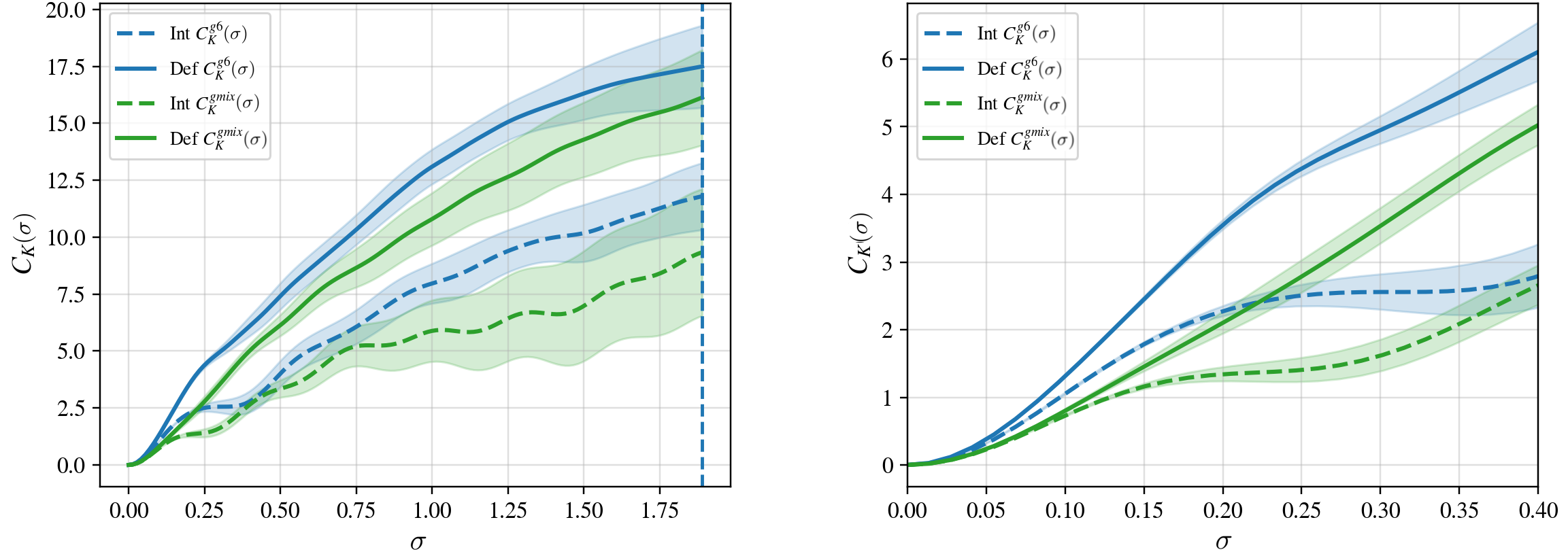}
    \caption{
Left: Krylov complexity $C_K^g(\sigma)$ for the seed observables $g_6$ and $g_{\rm mix}$, comparing the undeformed integrable flow ($\kappa=0$, labelled ``Int'') to the deformed flow at fixed $\kappa=0.9$ (labelled ``Def''). 
\textit{Right:} zoom into the small-$\sigma$ regime. 
The shaded bands indicate the spread over initial conditions (16--84\% quantiles), while the solid curves represent the corresponding median. The plots are generates for seeds with energies $H(\bullet~;\kappa=0)\in [1.44,\ 1.56]$.}
\label{fig:su3_ck_full_and_zoom}
\end{figure}

\begin{figure}[t]
    \centering
    \includegraphics[width=1.0\linewidth]{    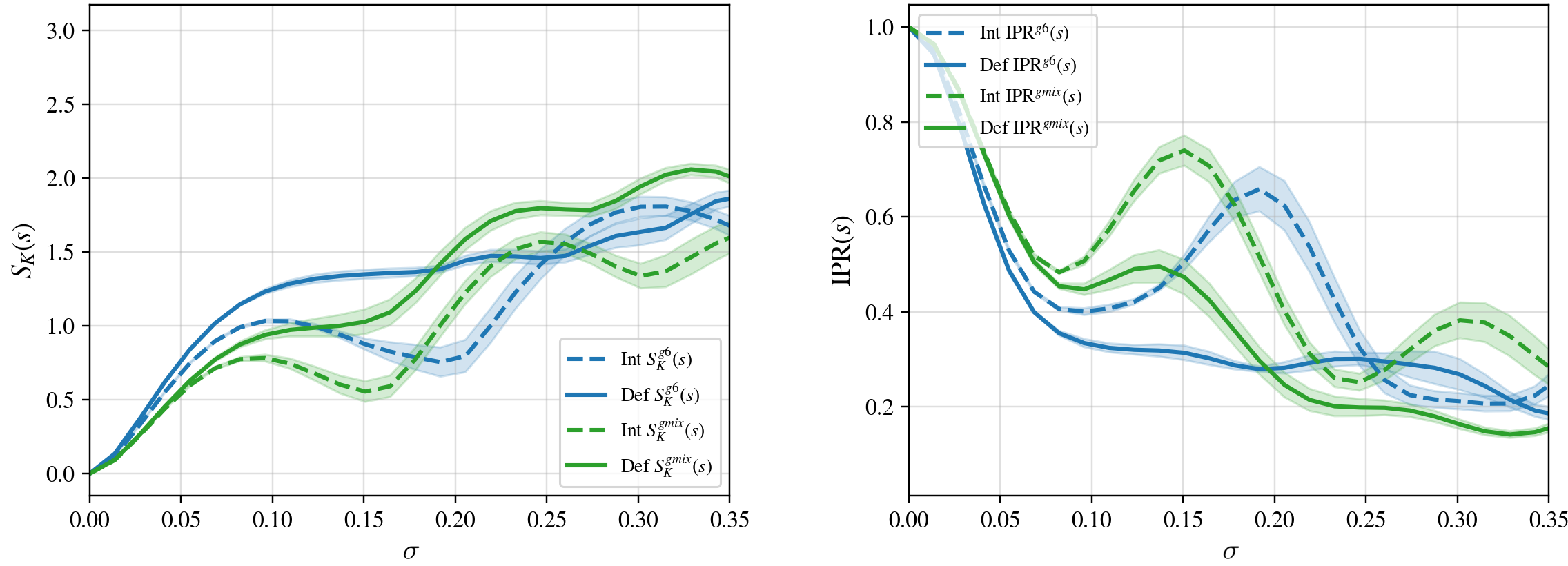}
\caption{
The Krylov Shannon entropy $S_K^g(\sigma)$ (left) and inverse participation ratio ${\rm IPR}^g(\sigma)$ (right) for the observables $g_6$ and $g_\mathrm{mix}$.
The integrable baseline ``Int'' ($\kappa=0$) exhibits relatively structured evolution, while the deformed system ``Def'' ($\kappa=0.9$) shows enhanced entropy production and reduced IPR at comparable $\sigma$, signalling a broader spread of amplitude over Krylov depth. The shading indicate the variations over the different seed initial conditions. The plots are generates for seeds with energies $H(\bullet~;\kappa=0)\in [1.44,\ 1.56]$.}
\label{fig:su3_entropy_and_ipr}
\end{figure}

\paragraph{Spread and delocalisation in Krylov space.}
We now quantify how the reduced complex Leigh-Strassler deformation induces observable-dependent redistribution of Koopman spectral weight, as detected by Krylov-space spreading and delocalisation. Fig.~\ref{fig:su3_ck_full_and_zoom} shows that switching on $\kappa$ enhances the rate of Krylov growth for both the sextic-angular probe $g_6$ and the momentum-angle mixing probe $g_{\rm mix}$.
This enhancement is visible already at small $\sigma$ (right panel), and persists throughout numerically-controlled range (left panel). Since the Koopman generator depends on $\kappa$ only through the deformation-induced non-separable sextic interaction in \eqref{eq:su3_Hdef}, the observed increase in $C_K^g(\sigma)$ is naturally interpreted as a signature of deformation-driven spectral transport: the observable ceases to remain supported on a sparse set of quasi-periodic Koopman frequencies and instead develops support on a denser set of near-resonant components.

Fig.~\ref{fig:su3_entropy_and_ipr} shows the entropy and IPR Krylov-diagnostics.
For $\kappa=0$ the wavepacket evolution remains comparatively coherent under the (deformed) Hamiltonian evolution: $S_K^g(\sigma)$ grows moderately and ${\rm IPR}^g(\sigma)$ remains relatively large, indicating that the Krylov amplitude is concentrated on a smaller effective range of depths.
When turning on $\kappa$, entropy production is enhanced while ${\rm IPR}^g(\sigma)$ decreases, consistent with delocalisation of the Krylov wavepacket over a broader set of Krylov levels. 

Finally, the relative ordering in terms of the resulting spread and delocalisation diagnostics  between observables is itself informative.  Across the same deformation strength, $g_{\rm mix}$ typically exhibits stronger delocalisation than the purely angular observable $g_6$, in the sense of larger $S_K^g(\sigma)$ and smaller ${\rm IPR}^g(\sigma)$ at comparable $\sigma$.  
This suggests  that $g_{\rm mix}$ couples more directly to deformation-activated momentum-angle mixing in the Hamiltonian flow, while $g_6$ is influenced by a more structured interplay between angular-sector redistribution and approximate constraints inherited from the near-integrable backbone. 
More precisely, Krylov delocalisation behaves as a sector-sensitive probe: it highlights which observables detect the earliest onset of resonance-mediated transport.

\begin{figure}[t]
    \centering
    \includegraphics[width=1.0\linewidth]{    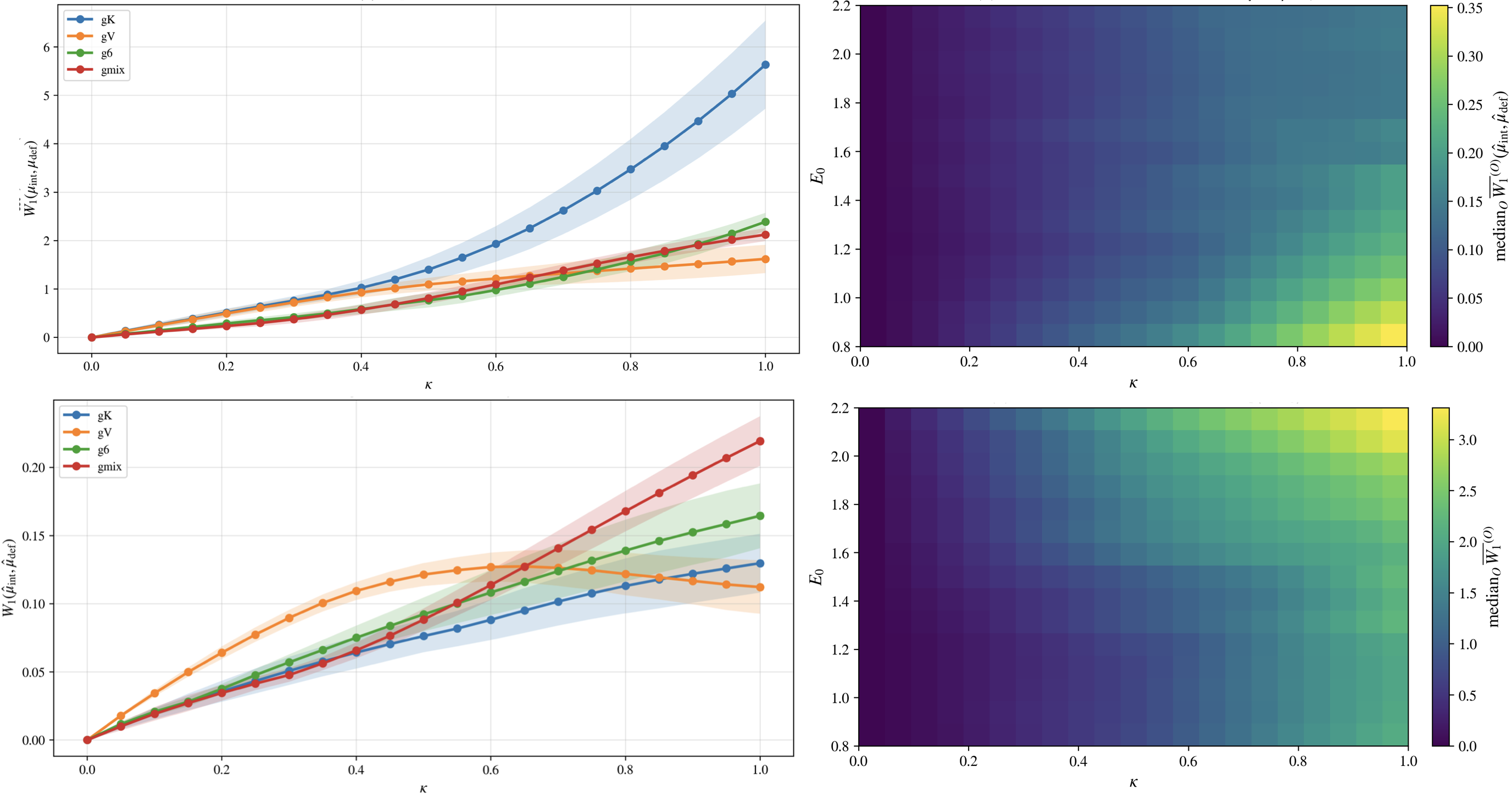}
    \caption{Energy-deformation sensitivity of spectral transport in the $SU(3)$ model.
Left panels: Wasserstein-1 distance between integrable and deformed Koopman spectral measures as a function of deformation $\kappa$ (curves show the observable dependence).
Right panels: heat maps of the median $W_1$ over observables as a function of energy shell $E_0$ and deformation.
Top: raw spectra, measuring absolute frequency transport.
Bottom: spectra normalised by mean and variance, isolating changes in spectral shape.
}
    \label{fig:T11_spectral_transportW1}
\end{figure}

\paragraph{Spectral transport.}
To quantify how integrability breaking reorganises the Koopman spectrum, we consider now the Wasserstein-1 distance between the spectral measures of the integrable and deformed flows, see eq.~\eqref{eq:wasserstein_int_def},
as a function of both deformation strength $\kappa$ and energy shell $E_0$ (evaluated as before from the undeformed Hamiltonian). This gives us in addition a quantitative probe for the energy-dependence of the spectral transport for increasing value of the deformation parameter. The resulting $W_1$-distances are shown in fig.~\ref{fig:T11_spectral_transportW1}. The upper panels use the raw spectral measures and therefore capture absolute transport in frequency space, while the lower panels compare spectra after normalising their mean and variance, isolating changes in spectral shape, rather than overall scale.

Note first that $W_1$ grows smoothly with $\kappa$ across the energy range, indicating a continuous redistribution of spectral weight away from the discrete integrable backbone. Second, the growth is strongly energy dependent: lower-energy shells exhibit enhanced sensitivity to the deformation, whereas higher energies remain closer to the integrable reference over a wider $\kappa$ interval. This energy hierarchy is consistent with a KAM-type picture in which invariant structures are progressively eroded, as well as the strong initial value dependence of the characteristic Lyapunov exponents \cite{McLoughlin:2022jyt} for this system.

Interpreted at finite Krylov resolution, the increase of $W_1$ signals the formation of dense frequency components from an initially sparse discrete spectrum. The normalised distance confirms that this effect is not merely a rigid shift or rescaling of frequencies but corresponds to genuine reshaping of the spectral measure. We therefore view the joint $(E_0,\kappa)$ dependence of $W_1$ as a quantitative proxy for the agglomeration into continuous components: it tracks how integrability breaking promotes the emergence of continuous spectral components and observable-dependent mixing in the Koopman spectrum.

\subsection{The near-Penrose limit of \texorpdfstring{$AdS_5 \times T^{1,1}$}{AdS5 x T{1,1}} solutions}\label{sec:T11}

We now turn to a different class of deformations of classical solutions which display a perturbation of a (reduced) integrable Hamiltotian perturbed by a integrability-breaking deformation.  A useful testing ground for weak integrability breaking in semiclassical string dynamics is provided by the near Penrose limit of $\text{AdS}_5\times T^{1,1}$. Picking a winding string ansatz, the worldsheet dynamics reduces to a finite-dimensional Hamiltonian system that is simple enough for explicit analysis but retains genuine near-integrable chaos. Depending on the Ansatz and as shown in \cite{Asano:2015qwa}, see also \cite{Basu:2011di,Ishii:2021asw,Panigrahi:2016zny}, corrections to the pp-wave background induce both integrable deformations and non-integrable deformations.  

These two deformations, the one preserving integrability and the one breaking it, is particularly in our goal to explore the signatures that weak classical chaos imprints on the Koopman-Krylov diagnostics defined in sec. \ref{sec:Krylovtoolkit}. Indeed, as we will see, the integrable deformation will display close to no change in the different diagnostics. In the integrability-breaking case, chaos does not translate into strong operator-space transport over Krylov subspace. Indeed Krylov spreads and inverse participation ratios are quantitatively but only weakly affected by the deformation. The dominant and robust signature of integrability breaking is instead captured by the spectral Wasserstein distance $W_1$, which responds directly to the reorganisation of the generator spectrum induced by the deformation. This provides an example of integrability breaking where, in the considered regime, chaos is primarily encoded at the spectral level rather than through global phase-space diffusion or fast operator growth. In particular, this highlights once again that weak integrability breaking in classical systems and in particular these semiclassical string solutions, contrary to the (strong) quantum chaos, does not translate into a universal signature in Krylov space. Instead, it provides a quantifiable transport in Krylov space generated by the deformed dynamics, where the model-dependence and observable-dependence is informative, signalling how the integrability breaking erodes the integrable phase space.

\subsubsection{A story of two motions: restricted $T^{1,1}$ and with AdS-direction}

We review the two deformations corresponding to reduced string motion in the near-Penrose limit of $\mathrm{AdS}_5\times T^{1,1}$ described in \cite{Asano:2015qwa}: a configuration including an AdS radial excitation, which remains integrable, and a motion restricted to the internal $T^{1,1}$ sector, where the deformation produces weakly non-integrable dynamics.

\paragraph{Motion with a radial component in AdS$_5$.}

We first consider the string motion Anstaz that includes a radial excitation in ${\rm AdS}_5$ in addition to a single internal $T^{1,1}$ mode, see \cite{Asano:2015qwa} for details and derivations. The reduced phase space is spanned by $(r,r_1,p_r,p_{r_1})$, and the Hamiltonian takes the form
\begin{equation}\label{eq:defH_T11}
H_{\mathrm{eff},\kappa} = H_0 + \kappa\, H_{\mathrm{int}}\,,
\end{equation}
with
\begin{equation}
H_0 = \frac12\bigl(p_r^2+p_{r_1}^2+r^2+(1+\alpha_1^2)r_1^2\bigr)\,,
\end{equation}
and interaction
\begin{equation}
H_{\mathrm{int}} = -\frac18 \left(p_r^2+p_{r_1}^2-r^2+(1+\alpha_1^2)r_1^2\right)^2 + \frac16 r^4 + \frac12(1-\alpha_1^2) r_1^4\,.
\end{equation}
This deformation, in the absence of any non-trivial Lyapunov exponents or irregular Poincar\'e sections is believed to preserve integrability.

To analyse this system we focus on two observables. We consider the quadratic ``size'' observable
\begin{equation}\label{def_T11_AdS_gQ}
g_Q(X)\equiv p_\rho^2+p_{r_1}^2+p_{r_2}^2+(1+\alpha_1^2)r_1^2+(1+\alpha_2^2)r_2^2\,,
\end{equation}
together with the internal mixing observable
\begin{equation}\label{def_T11_AdS_gmix}
g_{\rm mix}(X)\equiv r_1 p_{r_2}-r_2 p_{r_1}\,,
\end{equation}
which probes mode-rotation structure within the $T^{1,1}$ sector while remaining largely insensitive to the ${\rm AdS}$ radial excitation. 
Since the dynamics remains integrable for all $\kappa$, we expect close to no deformation dependence in the associated Krylov diagnostics, hence providing a controlled baseline against which the restricted $T^{1,1}$ case can be contrasted.

\paragraph{Motion restricted to $T^{1,1}$.}

A qualitatively different case is a string configuration whose dynamics is entirely restricted to the internal $T^{1,1}$ manifold, following the near-Penrose-limit analysis of \cite{Asano:2015qwa}. Wrapping the spatial worldsheet coordinate $\sigma$ along angles and retaining two transverse modes, the resulting light-cone Hamiltonian takes the form
\begin{equation}
H_{\kappa,\rm eff} = H_0 + \kappa\, H_\mathrm{int} \,,
\end{equation}
where the integrable part describes two decoupled harmonic oscillators,
\begin{equation}
H_0 = \frac12\Bigl(p_{r_1}^2 + p_{r_2}^2 + (1+\alpha_1^2) r_1^2 + (1+\alpha_2^2) r_2^2 \Bigr)\,,
\end{equation}
and the interaction is quartic,
\begin{equation}
H_\mathrm{int}= -\frac18\left(p_{r_1}^2+p_{r_2}^2+(1+\alpha_1^2)r_1^2 (1+\alpha_2^2)r_2^2\right)^2 -\frac12(\alpha_1 r_1^2-\alpha_2 r_2^2)^2 +\frac12(r_1^4+r_2^4)\,.
\end{equation}    
For $\kappa=0$ the system is integrable, while $\kappa\neq0$ introduces resonant couplings between the two modes that lead to the gradual destruction of invariant tori. As can be expected from the analysis revealing weak chaotic signatures in \cite{Asano:2015eha}, the deformation should not induce global chaotic motion. Even at finite $\kappa$, regular quasi-periodic trajectories coexist with chaotic ones, and instability is localised near resonant regions. As for the case before, transport is controlled by the progressive erosion of invariant tori rather than by large-scale mixing.

To probe this resonant structure we mirror the observables \eqref{def_T11_AdS_gQ} and \eqref{def_T11_AdS_gmix} defined for the previous anstaz, but now for phase-space for coordinates $X=(r_1,r_2,p_{r_1},p_{r_2})$. In particular, we consider the quadratic size observable
\begin{equation}
g_Q(X)\equiv p_{r_1}^2+p_{r_2}^2+(1+\alpha_1^2)r_1^2+(1+\alpha_2^2)r_2^2\,,
\end{equation}
together with the phase-space mixing observable $g_{\rm mix}$ with the same definition as in \eqref{def_T11_AdS_gmix}.

\begin{figure}[t]
    \centering
    \includegraphics[width=1.0\linewidth]{     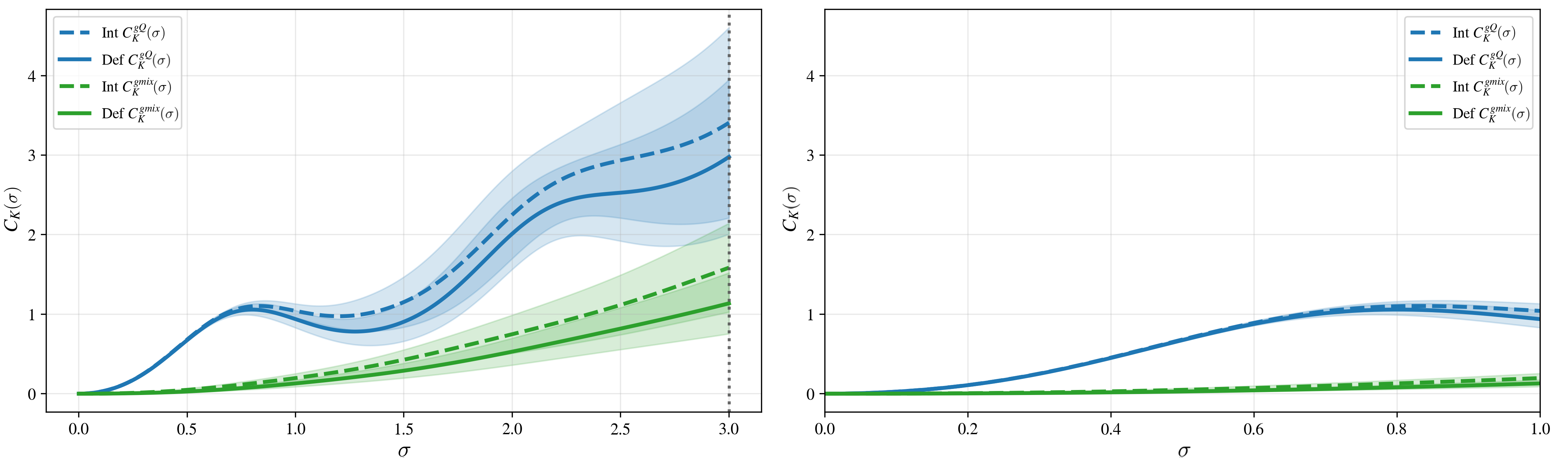}
    \caption{Left: Krylov spread $C_K(\sigma)$ for the integrable (Int) and deformed (Def) dynamics in the radial AdS embedding at $E=1.5$, $\kappa=0.170$, shown for $g_Q$ and $g_{\mathrm{mix}}$. Right: small-$\sigma$ zoom. Shaded bands indicate ensemble uncertainty.}
    \label{fig:T11_AdS_CK_FULL_AND_ZOOM__kappa_0p17}
\end{figure}

\subsubsection{Koopman-Krylov diagnostics for ${\rm AdS}$-radial system}\label{sec:T11AdSrad}

The integrable deformation of the $T^{1,1}$ background serves as a control example in which integrability is preserved and will help contrast the transport behavior in the integrability breaking case. Repeating the same observable-resolved comparison between the undeformed and deformed systems, as in the previous sections, we find that Krylov spread, delocalisation measures, and spectral transport remain nearly unchanged. Any differences are small and do not exhibit systematic growth with deformation strength. This near-rigidity confirms that the Koopman-Krylov diagnostics are sensitive specifically to integrability-breaking effects and do not generically produce artificial spectral drift under integrable deformations.

Although we choose two particular observables $g_Q$ and $g_\mathrm{mix}$, we have  scanned a substantially larger class of observables spanning the protected and mixing sectors. Across this broader scan the qualitative behaviour remains unchanged: in all cases the integrable deformation produces very small modifications of Krylov spread and spectral transport.

\paragraph{Spread and delocalisation in Krylov space.}
The Krylov spread in the original and integrably deformed systems is known in figure \ref{fig:T11_AdS_CK_FULL_AND_ZOOM__kappa_0p17}, showing that the deformation barely modifies the spread and delocalisation properties. Indeed, the Krylov complexity $C_K(\sigma)$ again does show non-trivial spread for both variables but the integrable and deformed curves display little difference. Indeed, whether a system is integrable or not, the unitary evolution will induce a non-trivial spread and delocalisation of the observable in the corresponding classical Hilbert space. However in this integrability preserving case, the deformation does not introduce any marked quantitative difference, highlighting that no chance of transport is occurring at the this level of the Krylov diagnostics.

\begin{figure}[t]
    \centering
    \includegraphics[width=1.0\linewidth]{     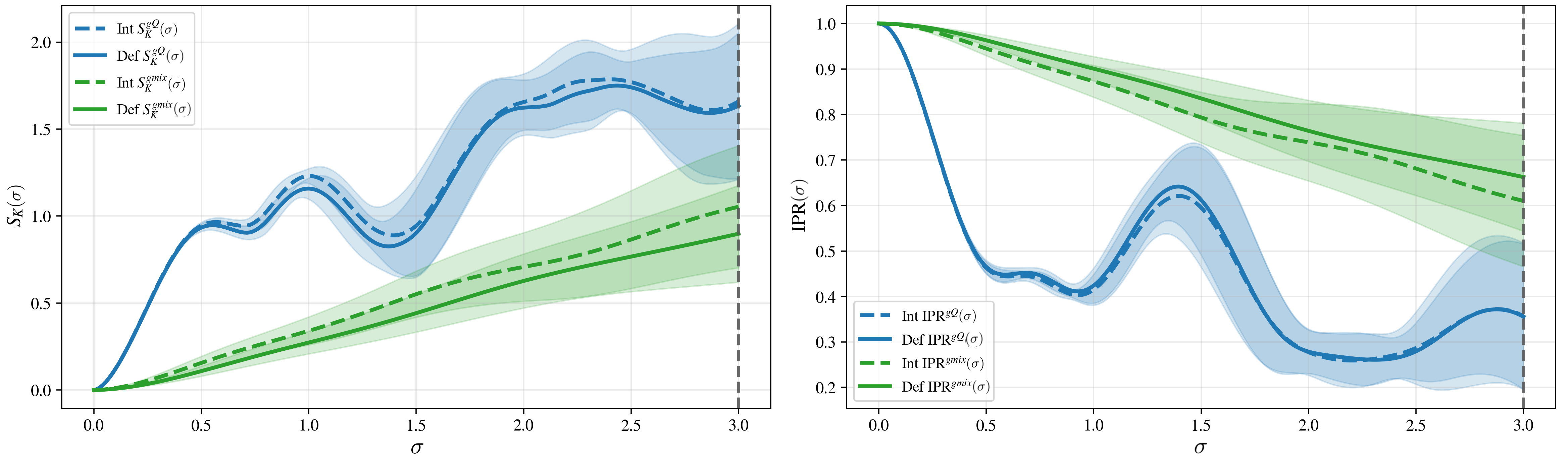}
    \caption{Krylov Shannon entropy $S_K(\sigma)$ (left) and inverse participation ratio $\mathrm{IPR}(\sigma)$ (right) for the integrable (Int) and deformed (Def) dynamics in the radial AdS embedding at $E=1.5$, $\kappa=0.170$, for $g_Q$ and $g_{\mathrm{mix}}$.}
    \label{fig:T11_AdS_ENTROPY_AND_IPR__kappa_0p17}
\end{figure}

The Krylov-delocalisation measures for this case are displayed in figure \ref{fig:T11_AdS_ENTROPY_AND_IPR__kappa_0p17}, and follow a similar trend as the Krylov spread. Again, the deformation does not change the delocalisation behaviour of the observables.  As before, hints again towards the  preservation of an integrable structure, i.e. the  foliation of the phase space in quasi-periodic orbits, through the absence of any substantial change in the motion on the  Krylov chain for a given observable.

\begin{figure}[t]
    \centering
    \includegraphics[width=1.0\linewidth]{     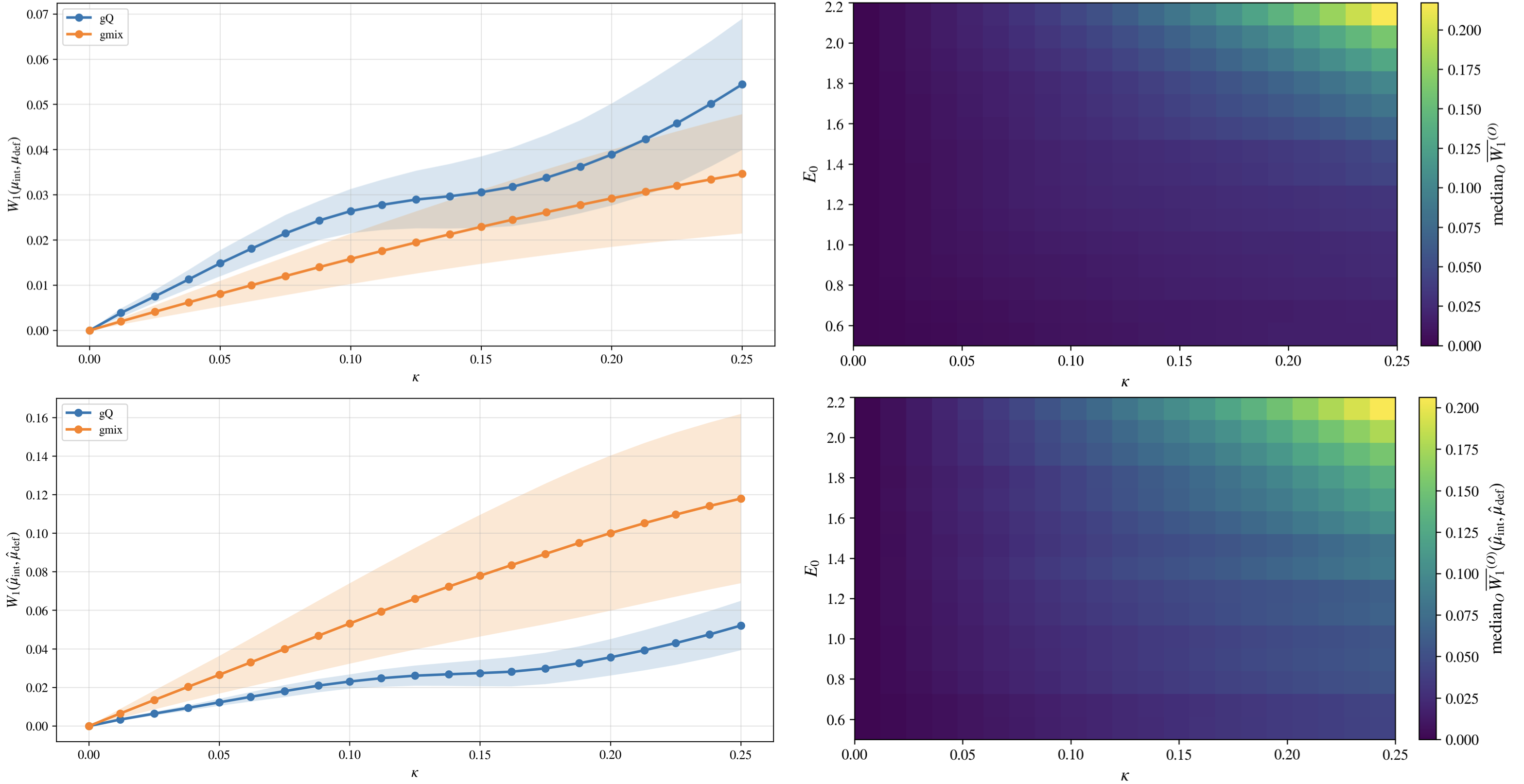}
    \caption{Energy-deformation sensitivity of spectral transport in the integrable $AdS\times T^{1,1}$ motion. Left panels: $W_1\!\left(\mu^{(\mathrm{int})}_g(E_0),\mu^{(\mathrm{def})}_g(E_0,\kappa)\right)$ versus $\kappa$ at fixed $E_0=1.350$ for representative probes $g_Q$ and $g_{\mathrm{mix}}$. Right panels: heat maps of the median $W_1$ over observables versus $(E_0,\kappa)$. Top: raw spectra (absolute frequency transport). Bottom: normalised spectra (shape transport).}
    \label{fig:T11_AdS_W1_energ}
\end{figure}

\paragraph{Spectral transport.}

Figure \ref{fig:T11_AdS_W1_energ} shows the Wasserstein-1 distance for the integrable $AdS\times T^{1,1}$ motion, resolved both in $\kappa$ at fixed energy $E_0$ (on the left) and across the $(E_0,\kappa)$ plane via the observable-median map (on the right right). The raw $W_1$ remains uniformly small throughout the scanned window, indicating that the deformation induces only a weak redistribution of Koopman spectral weight away from the integrable reference at the Krylov resolution used here. Comparing the raw and normalised distances, we find that the residual signal is not dominated by a uniform rigid drift of the spectrum; instead, any deformation effect is largely confined to mild reshaping of the spectral measure for certain probes, while remaining parametrically suppressed overall. In this sense the integrable deformation provides a ``null'' baseline: even though the Hamiltonian is deformed, the Koopman spectrum retains near-rigidity, and spectral transport is strongly constrained.

\subsubsection{Koopman-Krylov diagnostics in the restricted $T^{1,1}$ system}\label{sec:T11restricted}

For the restricted $T^{1,1}$ motion leading to weakly chaotic dynamics, the observable-resolved Koopman-Krylov comparison reveals a more pronounced reorganisation than in the spin-chain examples. This is particularly clear when juxtaposing the corresponding plots for the integrable deformation in the previous subsection. Indeed, here the deformation induces enhanced spread and delocalisation in Krylov space, but the magnitude of these effects varies strongly with the choice of observable. The most robust signal appears at the spectral level: Wasserstein distances between integrable and chaotic measures show significant transport, indicating substantial redistribution of Koopman frequencies. This behaviour is consistent with deformation-driven mixing concentrated in specific dynamical channels rather than uniform global chaos.

\begin{figure}[t]
    \centering
    \includegraphics[width=1.0\linewidth]{     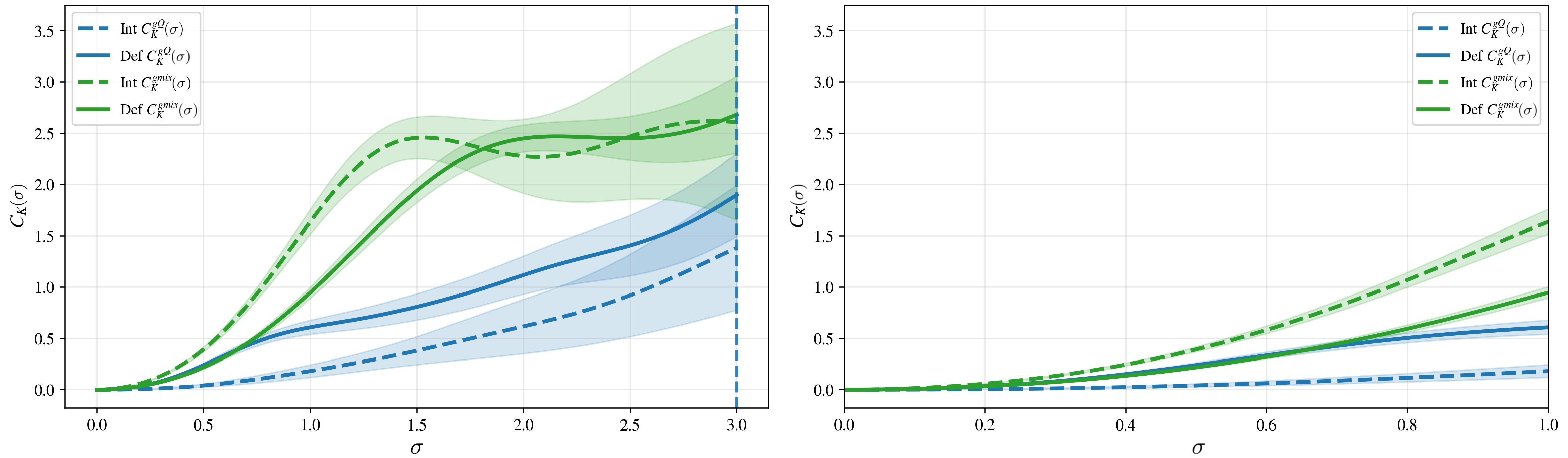}
    \caption{Left: Krylov spread $C_K(\sigma)$ for the integrable (Int) and deformed (Def) dynamics at $E=1.5$, $\kappa=0.170$, shown for $g_Q$ and $g_{\mathrm{mix}}$.  Right: small-$\sigma$ zoom. Shaded bands indicate ensemble uncertainty.}
    \label{fig:T11_restricted_CK_FULL_AND_ZOOM__kappa_0p17}
\end{figure}

\paragraph{Spread and delocalisation in Krylov space.}
Figure \ref{fig:T11_restricted_CK_FULL_AND_ZOOM__kappa_0p17} shows the Krylov spread for the restricted $T^{1,1}$ system at fixed energy $E=1.5$ and deformation strength $\kappa=0.17$, where the deformation modifies the effective worldsheet Hamiltonian without activating an additional dynamical direction. 
In this setting, the Krylov complexity $C_K(\sigma)$ exhibits a clear change in transport under deformation for both observables. The $g_Q$-observable shows a mild but robust increase in the deformed theory relative to the integrable baseline, while the mixed probe $g_{\mathrm{mix}}$  grows substantially faster already at intermediate $\sigma$, reaching markedly larger Krylov depth compared to the undeformed case. At small $\sigma$ (zoom panel in fig.~\ref{fig:T11_restricted_CK_FULL_AND_ZOOM__kappa_0p17}), the hierarchy between probes is already visible: $C^{g^{\mathrm{mix}}}_K(\sigma)$ rises parametrically faster than $C^{g_Q}_K(\sigma)$, and the difference between integrable and deformed curves is most pronounced for the mixed observable. This indicates that the deformation produces a genuine reorganisation of Koopman spectral weight, but the effect remains strongly observable dependent: in the restricted system, only probes with sufficient overlap with deformation-sensitive channels display strong Krylov drift.

\begin{figure}[t]
    \centering
    \includegraphics[width=1.0\linewidth]{     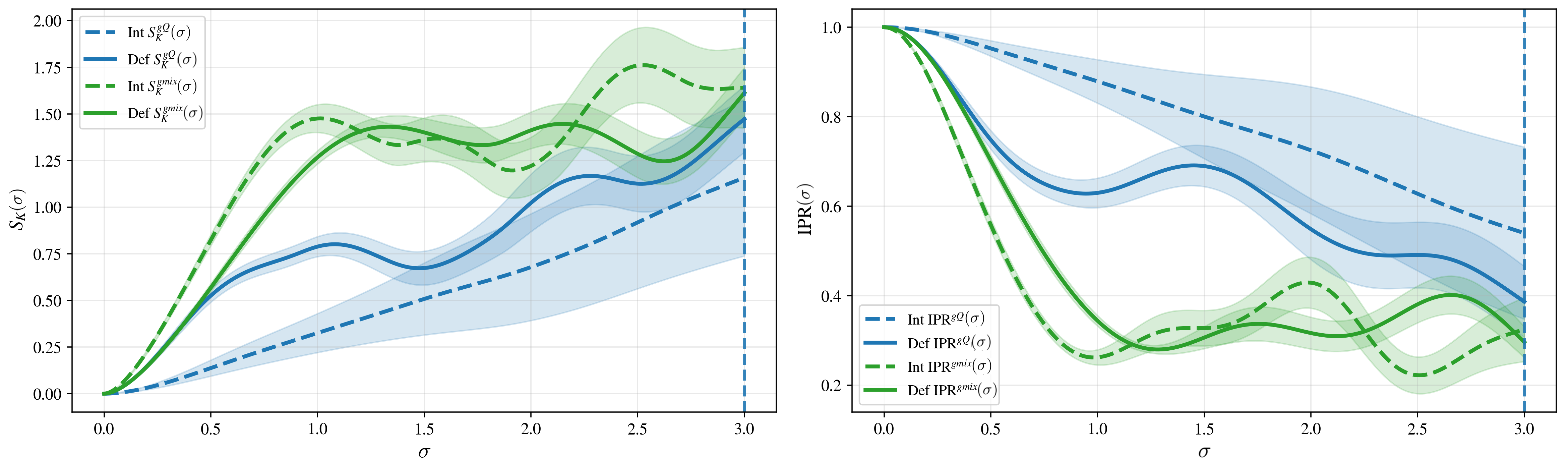}
    \caption{Krylov Shannon entropy $S_K(\sigma)$ (left) and inverse participation ratio $\mathrm{IPR}(\sigma)$ (right) for the integrable (Int) and deformed (Def) dynamics at $E=1.5$, $\kappa=0.170$, for $g_Q$ and $g_{\mathrm{mix}}$. Shaded bands indicate ensemble uncertainty.}
    \label{fig:T11_restricted_ENTROPY_AND_IPR__kappa_0p17}
\end{figure}

To assess whether the enhanced Krylov spread is associated with delocalisation, we turn to the Krylov Shannon entropy $S_K(\sigma)$ and inverse participation ratio $\mathrm{IPR}(\sigma)$ in figure \ref{fig:T11_restricted_ENTROPY_AND_IPR__kappa_0p17}. In the restricted $T^{1,1}$ system, both diagnostics point to a deformation-induced spreading of the Krylov wavefunction, but again with a strong observable dependence. For the $g_Q$ probe, the entropy displays a small $\sigma$ rise followed by a oscillatory behaviour, suggesting that Krylov weight populates a moderate band of Lanczos directions while retaining substantial structure. Correspondingly, the IPR exhibits a pronounced early drop and subsequent fluctuation. By contrast, the mixed probe $g_{\mathrm{mix}}$  shows a more steady entropy increase and a smoother overall IPR suppression, signalling broader occupation of Krylov levels. 

In short, we see observable dependent change in transport, signalling that certain directions in phase space as seen by the chosen observables are perturbed due to the particular form of the deformation. Importantly, as we choose the analogous observable  as for the for the integrability-preserving deformation, this suggests that this transport is associated to the slow and gradual destruction of the integrable structure of the phase space.

\paragraph{Spectral transport.}

\begin{figure}[t]
    \centering
    \includegraphics[width=1.0\linewidth]{     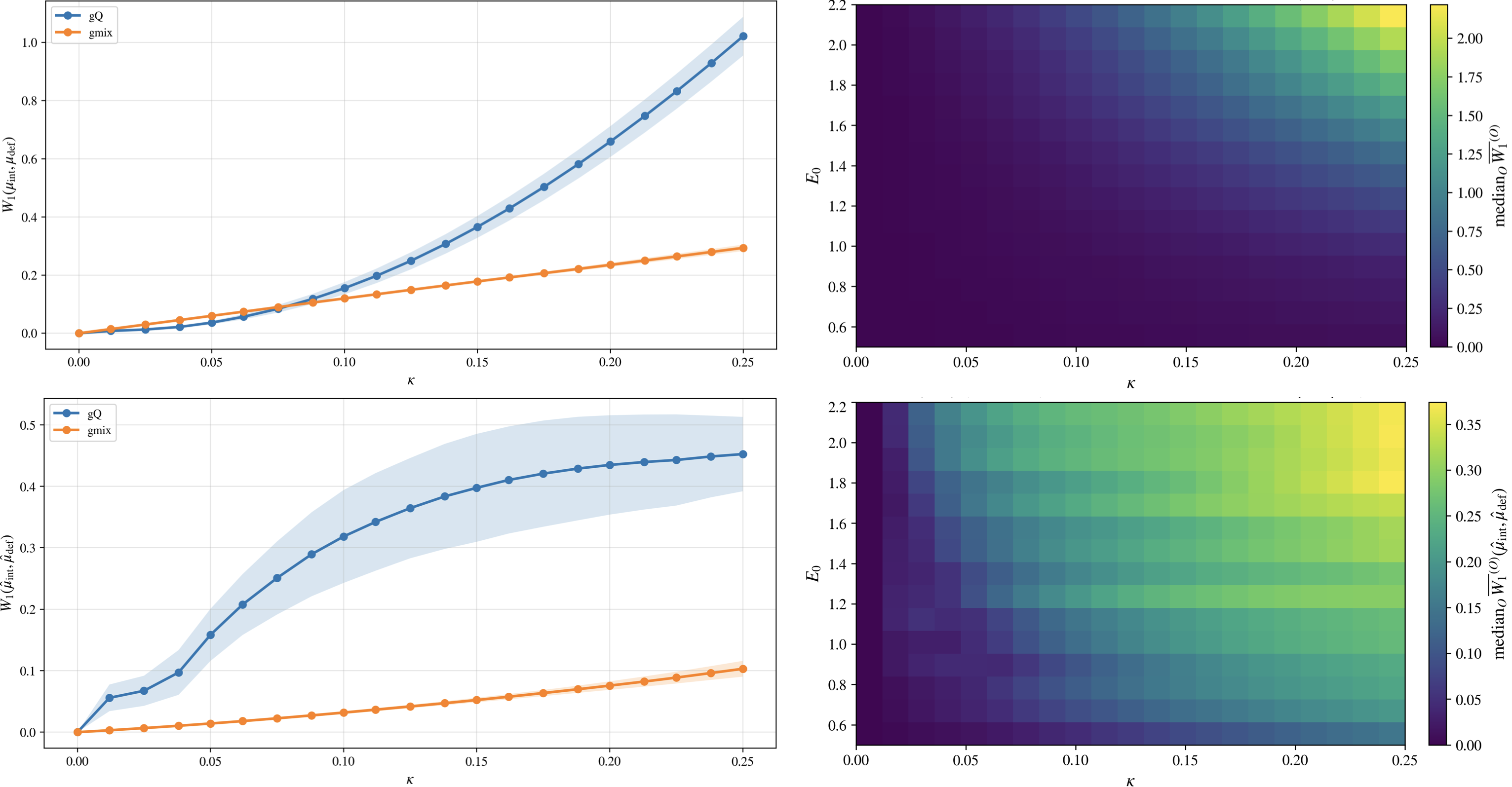}
    \caption{Energy-deformation sensitivity of spectral transport in the restricted $T^{1,1}$ dynamics. Left panels: $W_1$ versus $\kappa$ at fixed $E_0=1.350$ for $g_Q$ and $g_{\mathrm{mix}}$, illustrating strong observable dependence. Right panels: heat maps of the median $W_1$ over observables versus $(E_0,\kappa)$. Top: raw spectra (absolute transport). Bottom: normalised spectra (shape transport). }
    \label{fig:T11_restricted_W1_energy}
\end{figure}

In contrast to the transport shown in fig.~\ref{fig:T11_AdS_W1_energ} induced in the integrable deformation, figure \ref{fig:T11_restricted_W1_energy} demonstrates that the restricted $T^{1,1}$ dynamics exhibits a pronounced deformation-driven reorganisation of the Koopman spectrum as quantified by $W_1$, as defined in eq.~\eqref{eq:wasserstein_int_def}. At fixed energy $E_0$, the growth of $W_1$ with $\kappa$ is strongly observable dependent: certain observables undergo substantial spectral transport while others remain comparatively inert. This observable selectivity persists in the $(E_0,\kappa)$ maps (right panels), where the median over observables increases with $\kappa$ and is non-uniform in energy. The distinction between the raw and normalised $W_1$ (see discussion around eq. \eqref{eq:W1normalised} for the definitions) further shows that the effect is not reducible to a simple shift or rescaling of frequencies: a sizeable component survives normalisation, signalling genuine reshaping of the spectral measure (redistribution of weight between frequency packets) rather than purely rigid transport. We interpret this as evidence that the integrability-breaking interaction couples efficiently only through specific resonance channels, i.e. the deformation is ``felt'' only on particular regions of phase space and for observables with sufficient overlap with those channels; this picture is corroborated by the perturbative analysis presented in the appendix \ref{app:perturb}.

\section{Conclusion}\label{sec:conclusions}

In this work we formulated a classical version of Krylov methods, which was originally developed in quantum many-body systems, by exploiting the Koopman-von Neumann formulation of classical mechanics, and apply it to integrability breaking in semiclassical string dynamics. The key idea is to recast classical Hamiltonian flow as linear evolution on a Hilbert space of observables and to use Krylov constructions to probe how the spectral content of selected observables reorganises under the dynamics. This yields a concrete pipeline based on Koopman evolution, gEDMD approximation, and Lanczos tridiagonalisation that is naturally suited to classical systems, and in particular to the weakly non-integrable regimes arising in semiclassical string reductions. We applied this construction to the two-loop-deformed $SU(2)$ and Leigh-Strassler-deformed $SU(3)$ spin-chain sectors, as well as to near-Penrose limits of strings on $\mathrm{AdS}_5 \times T^{1,1}$, we computed observable Krylov spreads (Krylov complexity, inverse participation ratios and entropies), spectral-measure deformations, and phase-space filamentation diagnostics. These quantities resolve how weak deformations redistribute observable weight and reorganise classical dynamics.

Importantly, while the formal construction applies once the relevant subspace and hopping approximation are identified, this does not imply that Krylov spreading provides a uniform characterisation of chaos across parameter space. In weakly non-integrable regimes, the response does not exhibit the universal ballistic growth characteristic of strongly chaotic quantum systems.

The motivation for this approach is that conventional diagnostics of classical chaos often provide binary indicators and do not necessarily resolve how dynamical structures reorganise across parameter space, especially in high-dimensional systems. By contrast, Krylov methods track how different observables reorganise under deformation. In weakly chaotic classical systems, spreading is strongly observable dependent and tied to specific resonant channels rather than exhibiting the universal ballistic growth familiar from quantum many-body settings. The visibility of Krylov signatures therefore depends sensitively on the chosen observable and projection, and should not be interpreted as a model-independent scalar measure of chaos. This sensitivity allows the Koopman-Krylov framework to resolve distinct mechanisms of integrability breaking and to map deformation-induced spectral transport across phase space.

Across all systems studied in this work, the response depends strongly on the observable and on the underlying resonance structure. In particular, the onset of chaos is not marked by a sharp or universal behaviour in Krylov complexity, but by gradual and resonance-specific redistribution of spectral weight. Instead, the deformation induces controlled and quantifiable transport whose magnitude and structure are model- and observable-dependent. Integrability breaking manifests through resonance-driven delocalisation in Krylov space, phase-space filamentation, and spectral reshaping, with the detailed pattern encoding how the integrable phase-space foliation is eroded. 

The present work should be viewed as a first exploration of this observable Koopman-Krylov approach in semiclassical string dynamics. While the framework provides a systematic way to analyse spectral transport and observable spreading, many aspects remain to be understood, including the optimal choice of observables for a given system, the robustness of diagnostics under changes of dictionary and sampling measure, and the relation between different Krylov probes and specific dynamical mechanisms. In particular, the framework highlights that Krylov diagnostics in classical systems are intrinsically definition dependent: different observables probe different phase-space structures. Rather than a drawback, this feature encodes detailed information about which invariant structures are being eroded and how transport unfolds. Further developments along these directions should clarify the range of applicability and full diagnostic power of observable-based Krylov methods for classical and near-integrable dynamics.

\vspace{20pt}

\noindent
We conclude with a number of interesting future perspectives.

\vspace{-3pt}
\paragraph{Application to LLM microstate geometries and trapping.}
The present work establishes a general Koopman-Krylov framework in a class of non-trivial semiclassical string systems, but many aspects of the Koopman-Krylov framework remain to be explored. Because the diagnostics are inherently system- and observable-dependent, a natural next step is to apply them to other settings where conventional chaos probes face limitations. A particularly promising arena is provided by LLM microstate geometries, where integrability breaking is closely tied to trapping phenomena and chaotic scattering \cite{Berenstein:2025ese,Bena:2017upb,Berenstein:2023vtd}. In these backgrounds, standard tools such as Poincaré sections and Lyapunov exponents are difficult to interpret due to the high dimensionality of phase space. The Koopman-Krylov framework offers a complementary strategy: by selecting probes tailored to specific dynamical channels, one may isolate and quantify trapping and transport mechanisms that are otherwise obscured.

\paragraph{Universality in deformation-classes ?}
More broadly, our results raise the question of whether restricted forms of universality may emerge within classes of integrability breaking. While no universal Krylov signature is expected across arbitrary systems, it is conceivable that deformations sharing a common physical origin  (such as higher-loop corrections or geometrically induced trapping) exhibit characteristic patterns of observable spreading and spectral transport. Identifying such patterns would provide a finer classification of weak chaos in string dynamics and clarify how different mechanisms of integrability breaking organise phase-space transport.

\paragraph{Koopman-Krylov diagnostics in pseudointegrable billiards.} 
One future direction is to extend the Koopman-Krylov framework to pseudointegrable billiards, i.e., polygonal billiards with rational internal angles, where classical trajectories live on compact invariant surfaces whose topology is generically not a torus but a multiply handled sphere genus $g>1$ \cite{RICHENS1981495,Jain_2017}. These systems provide a controlled bridge between integrable genus one torus foliated phase space motion and the richer structures available on higher-genus manifolds, and they offer a concrete setting to test whether observable-resolved Koopman evolution can probe the topology of the classical phase space dynamics \cite{RICHENS1981495,Jain_2017,Balasubramanian:2024ghv}. In particular, one could consider families of billiards in which the genus is tunable and then study how observable-induced spectral measures and Krylov delocalisation reorganise as $g$ changes. In this way, Koopman-Krylov diagnostics could provide a quantitative handle on how phase-space topology constrains spectral redistribution and mixing pathways in near-integrable dynamics.

\vspace{30pt}

\subsection*{Acknowledgments}
We thank Ofer Aharony, Vijay Balasubramanian, Pawel Caputa, Tristan McLoughlin for very interesting discussions.  RND is supported by the PRIME programme of the German Academic Exchange Service (DAAD), with funds from the German Federal Ministry of Research, Technology and Space (BMFTR) and is also supported by Germany's Excellence Strategy through the W\"urzburg-Dresden Cluster of Excellence ctd.qmat - Complexity, Topology and Dynamics in Quantum Matter (EXC 2147, project-id 390858490), and by the Deutsche Forschungs- gemeinschaft (DFG) through the Collaborative Research centre ``ToCoTronics'', Project-ID 258499086-SFB 1170. SD would like to thank the participants and organisers of Iberian Strings 2026 for interesting discussions and the opportunity of the presenting preliminary results of this work.

\appendix
\section{Perturbative analysis of KAM breaking and resonances}\label{app:perturb}
In this appendix we perform a linear perturbative KAM-type analysis of how the deformations considered in the main text destabilise invariant tori and generate resonant transport.  In particular,  we first expands the Hamiltonian around a stable equilibrium and brings the quadratic part to a set of decoupled harmonic oscillators.  The resulting action-angle variables provide a natural coordinate system in which the unperturbed dynamics is integrable and organised by invariant tori.  The leading nonlinear corrections can then be written as a finite Fourier series in the angles.  Each Fourier harmonic selects a potential resonance surface defined by an integer relation $k\cdot \Omega \simeq 0$ among the linear frequencies.  Near such a surface the dynamics reduces to an effective pendulum-like resonant Hamiltonian, whose amplitude controls the width of the corresponding chaotic layer in phase space. See for example \cite{de2001tutorial} for a pedagogical and practical guide to KAM theory and e.g. \cite{ferraz2007canonical,Cincotta_2014} for explicit applications.

Our goal is not to construct a full KAM proof, but to identify which resonant channels are activated by the string deformations and how strongly they couple to the dynamics.  This perturbative information will allow us to interpret the structure seen in the Koopman-Krylov diagnostics: tuning the deformation parameters moves the system across resonant surfaces in action space, while the strength of the nonlinear couplings sets the thickness of the associated transport layers.  We first apply this mechanism in the rigid $SU(2)$ sector and then adapt the same reasoning to the near-Penrose $T^{1,1}$ systems.

\subsection*{LL-limit two-loop SU(2)-sector}\label{sec:su2_action_angle_resonance}

For the rigid $SU(2)$ circular-ansatz sector reviewed in sec.~\ref{sec:koopmansu2}  we can perturbatively analyse the near-integrable regime. Indeed, one can put the two-loop Hamiltonian \eqref{eq:Ham_su2_ch} into the so-called Birkhoff normal form and read off, perturbatively, which invariant tori are most susceptible to destabilisation driven by the deformation parameters.

To do so, we expand around the stable equilibrium $q=0$ of the pendulum potential and work in the Ostrogradsky phase space $(q,v;p_0,p_1)$ with $v=q^{(1)}$. Using \eqref{eq:Ham_su2_ch} and $\cos q = 1-\tfrac12 q^2 + \tfrac1{24}q^4+\cdots$, the Hamiltonian decomposes as
\begin{equation}
H = H_2 + H_4 + O(6)\,,
\end{equation}
with quadratic and quartic part
\begin{equation}\label{eq:H4_KAM_perturb}
H_2 = v\,p_0 + \frac{p_1^2}{4|a_1|} + a_0 v^2 + \frac{\omega}{2} q^2\,,\quad H_4 = (a_1+a_2)\,v^4 - \frac{\omega}{24}\,q^4\,,\qquad (a_1<0)\,,
\end{equation}
The linearised equation of motion the follows from $H_2$ (equivalently by \eqref{eq:def_circul_string_eoms} at small amplitude) has two characteristic frequencies $\Omega_\pm$ determined by
\begin{equation}
-a_1 \Omega^4 - a_0 \Omega^2 + \frac{\omega}{2}=0\,,\quad \text{i.e.}\quad  \Omega_\pm^2 = \frac{-a_0 \pm \sqrt{a_0^2 + 2 a_1 \omega}}{-2a_1}\,.
\end{equation}
In particular, for small $|a_1|$ one finds
\begin{equation}
\Omega_-^2\simeq\frac{\omega}{2a_0}\,, \qquad \Omega_+^2\simeq\frac{a_0}{|a_1|}\,,
\end{equation}
so that increasing $|a_1|$ lowers the fast frequency $\Omega_+$ and continuously sweeps the ratio $\Omega_-/\Omega_+$ through a dense set of low-order rational values.  It is therefore $a_1$ that controls the generation and location of resonant surfaces in action space.

There exists a linear symplectic transformation bringing $H_2$ to two decoupled harmonic oscillators,
\begin{equation}
H_2 = \Omega_- I_- + \Omega_+ I_+\,,
\end{equation}
with actionangle variables $(I_\pm,\theta_\pm)$ defined by
\begin{equation}\label{eq:AA_su2}
Q_\pm = \sqrt{\frac{2I_\pm}{\Omega_\pm}}\cos\theta_\pm\,\quad P_\pm = -\sqrt{2I_\pm \Omega_\pm}\sin\theta_\pm\,, \qquad \theta_\pm\sim\theta_\pm+2\pi\,.
\end{equation}
We will not need the explicit linear map $(q,v;p_0,p_1)\mapsto(Q_\pm,P_\pm)$ below; it can be obtained by diagonalising the quadratic form $H_2$ as a Hamiltonian matrix problem.  What matters for resonance theory is that $q$ and $v=q^{(1)}$ are linear combinations of $Q_\pm$ and $P_\pm$ with coefficients fixed by $a_0,a_1,\omega$ and hence ultimately by $\Omega_\pm$.

To leading order in the normal-form variables one may write
\begin{equation}
q = \alpha_- Q_- + \alpha_+ Q_+\,,\quad v = \beta_- P_- + \beta_+ P_+\,,
\end{equation}
with action-angle variables given in \eqref{eq:AA_su2} and for some real constants $\alpha_\pm,\beta_\pm$ determined by the linear symplectic transformation (and thus by $a_0,a_1,\omega$).  
Inserting the action-angle parametrisation for $(Q_\pm,P_\pm)$ yields 
\begin{align}\label{eq:su2_perturb_qv_AA}
q &= \alpha_- \sqrt{\frac{2I_-}{\Omega_-}}\cos\theta_- + \alpha_+ \sqrt{\frac{2I_+}{\Omega_+}}\cos\theta_+\,,\\
v &= -\beta_- \sqrt{2I_- \Omega_-}\sin\theta_- - \beta_+ \sqrt{2I_+ \Omega_+}\sin\theta_+\,.
\end{align}
The quartic perturbation $H_4=(a_1+a_2)v^4-\frac{\omega}{24}q^4$ therefore becomes an explicit finite Fourier series in the angles,
\begin{equation}\label{eq:fourier_decomp_su2_H4}
H_4(I,\theta)=\sum_{k\in\mathbb Z^2,\ |k_1|+|k_2|\le 4} h_k(I)\,e^{i(k_1\theta_-+k_2\theta_+)}\,,
\end{equation}
with coefficients polynomial in $I_\pm$. For example, the $v^4$ term alone generates the harmonics
\begin{equation}\label{eq:modessu2}
k\in\{(\pm 4,0), (0,\pm 4), (\pm 2,0), (0,\pm 2), (\pm 2,\pm 2), (\pm 2,\mp 2)\}\,,
\end{equation}
together with an angle-independent part.

Going back to the leading order contribution to the integrable deformation of the Hamiltonian
\begin{equation}
H(I,\theta) = \Omega_- I_- + \Omega_+ I_+ + \varepsilon H_4(I,\theta) + O(6)\,, \qquad \varepsilon\sim \frac{\lambda}{L^2}\,,
\end{equation}
the small denominators controlling canonical perturbation theory are
\begin{equation}
k \cdot\Omega = k_1\Omega_- + k_2\Omega_+\,,\qquad k\in\mathbb Z^2\setminus\{0\}\,.
\end{equation}
As reviewed in sec. \ref{sec:weak_chaos_koopman}, a torus with actions $I=(I_-,I_+)$ is susceptible to destabilisation when there exists a low-order harmonic $k$ present in $H_4$ such that $|k\cdot\Omega|$ is small compared to the corresponding coupling. In the present case $\frac{\Omega_-}{\Omega_+}\simeq\sqrt{\frac{\omega |a_1|}{2 a_0^2}}$, and as $a_1$ increases it visits potential Fourier modes of the interaction Hamiltonian\footnote{For example the first that is non-trivial and that is hit when $a_1$ increases is, by inspecting the list in \eqref{eq:modessu2}, $k=(2,-2)$ or $k=(-2,2)$, i.e. for which $\Omega_- \approx \Omega_+\,$.}.

Performing a near-identity canonical transformation to remove all non-resonant Fourier modes and introducing resonant coordinates $\psi=k\cdot\theta$ with conjugate action $J$, one obtains after averaging over the fast angle the standard resonant normal form
\begin{equation}\label{eq:resH}
    H_{\mathrm{res}}=k\cdot\Omega\,J+\varepsilon |h_k(I)|\cos(k\cdot\theta)\,,
\end{equation}
i.e. the Hamiltonian of a pendulum governing the slow dynamics near the resonant torus, see e.g. \cite{arnold1989mathematical}. In \eqref{eq:resH}, $J$ is the resonant action and $h_k(I)$ is the Fourier coefficient of $H_4$ for that harmonic as defined in eq.~\eqref{eq:fourier_decomp_su2_H4}. One can compute by plugging \eqref{eq:su2_perturb_qv_AA} directly into the action \eqref{eq:H4_KAM_perturb} and identifying the amplitude $h_k$ for the $k=(2,-2)$-mode in \eqref{eq:fourier_decomp_su2_H4} to be 
\begin{equation}
    |h_{(2,-2)}| \propto \left| (a_1+a_2)\,\beta_-^2\beta_+^2 - \frac{\omega}{24}\alpha_-^2\alpha_+^2 \right|\,.
\end{equation}
Taken together we see that the deformation parameter $a_1$ controls the location of resonant surfaces by tuning the ratio of the linear frequencies $\Omega_-/\Omega_+$, while the quartic couplings, controlled by $(a_1+a_2)$, set the amplitude of the resonant Fourier modes and hence the width of the corresponding resonant layers in phase space.

This provides a linear-order analysis supporting the pattern seen in the Krylov diagnostics plotted in fig.~\ref{fig:su2_W1_heatmap_median}. The heatmaps provide a direct evidence that integrability breaking in this sector is resonance-driven rather than uniformly chaotic: tuning $a_1$ moves the system across resonant surfaces, while $a_2$ controls the strength of mixing within those resonant layers.

\subsection*{Near-Penrose $T^{1,1}$ systems}
\label{sec:T11_action_angle_resonance}

We now adapt the normal-form KAM analysis to the specific near-Penrose $T^{1,1}$ models and observables used in sec.~\ref{sec:T11}.
The goal is not to re-diagnose chaos, but to isolate (i) which slow angles can arise under the deformation and (ii) which of our Koopman probes have parametrically large overlap with the corresponding resonant channels.

A resonance in a two-degree-of-freedom integrable system arises when the frequencies satisfy an approximate integer relation $k_1\omega_1+k_2\omega_2\simeq0$ \cite{duistermaat1980global,arnold1989mathematical}, which produces a slow angle and small denominators in perturbation theory. The simplest nontrivial case is the $1{:}1$ resonance, $\omega_1\simeq\omega_2$, for which the relative phase $\psi=\phi_1-\phi_2$ evolves slowly while the orthogonal combination $\chi=\phi_1+\phi_2$ remains fast. In this regime the quadratic Hamiltonian depends only on the total action $I_1+I_2$, leading to an approximate $S^1$ symmetry and an effective one-degree-of-freedom dynamics for the slow phase $\psi$.

For weakly coupled oscillators with quartic interactions, the perturbation generically produces Fourier components $\cos(2\phi_i)$ and $\cos(2(\phi_1\pm\phi_2))$, so the only internal small-denominator condition supported at leading order is the $1{:}1$ resonance $\omega_1\simeq\omega_2$. In the restricted near-Penrose $T^{1,1}$ model the interaction explicitly contains the resonant harmonic $\cos(2(\phi_1-\phi_2))$, while no additional mixed harmonics are generated, placing the system in the universal class of weakly coupled oscillators near a single symmetry-selected $1{:}1$ resonance.
\paragraph{Restricted $T^{1,1}$ system and $1{:}1$ resonance.}
In the restricted near-Penrose model the Hamiltonian, as before, we write the deformation as $H = H_0 + \lambda H_{\rm int}$, with unperturbed part now grouping the terms as follows
\begin{equation}
H_0=\frac12\Bigl(p_{r_1}^2 + p_{r_2}^2 + \omega_1^2 r_1^2 + \omega_2^2 r_2^2 \Bigr)\,,\qquad \omega_i^2 = 1 + \alpha_i^2\,,
\end{equation}
and quartic interaction
\begin{equation}
H_\mathrm{int}=-\frac18 Q^2-\frac12 D^2+\frac12\left(r_1^4 + r_2^4\right)\,,
\end{equation}
\begin{equation}
Q = p_{r_1}^2 + p_{r_2}^2 + \omega_1^2 r_1^2 + \omega_2^2 r_2^2\,,\qquad D = \alpha_1 r_1^2 - \alpha_2 r_2^2\,.
\end{equation}
Introducing action-angle variables $(I_i,\phi_i)$ for each oscillator,
\begin{equation}
r_i = \sqrt{\frac{2I_i}{\omega_i}} \sin\phi_i,\qquad
p_{r_i} = \sqrt{2I_i\omega_i}\cos\phi_i,
\end{equation}
the quadratic Hamiltonian and second  are angle-independent
\begin{equation}
H_0 = \omega_1 I_1 + \omega_2 I_2\,,\quad Q = 2(\omega_1 I_1 + \omega_2 I_2)\,.
\end{equation}
The term $-\tfrac18 Q^2$ therefore produces only action-dependent corrections (nonlinear frequency shifts) and does not generate Fourier harmonics or resonant couplings at order $\lambda$.

The angle dependence of $H_{\rm int}$ arises solely from $r_i^4$ and $D^2$, whose trigonometric reduction yields a finite Fourier spectrum consisting of a constant component and the harmonics
\begin{equation}
\cos(2\phi_i),\qquad \cos(4\phi_i),\qquad \cos\bigl(2(\phi_1\pm\phi_2)\bigr)\,,
\end{equation}
with coefficients polynomial in the actions. Among the mixed combinations, the only candidate for a slow phase is $\psi := \phi_1-\phi_2$, since the interaction contains a term proportional to $\cos(2\psi)$.

To make the resonance structure explicit, introduce fast and slow angles
\begin{equation}
\psi=\phi_1-\phi_2,\qquad \chi=\phi_1+\phi_2\,,
\end{equation}
with conjugate actions $I_\pm=\tfrac12(I_1\pm I_2)$. The unperturbed frequencies are
\begin{equation}
\dot\psi=\omega_1-\omega_2,\qquad \dot\chi=\omega_1+\omega_2\,.
\end{equation}
For positive $\omega_i$, the combination $\chi$ remains fast, whereas $\psi$ becomes slow near the $1{:}1$ surface $\omega_1\simeq\omega_2$. At this perturbative order and within the restricted truncation, no additional independent slow combinations arise, so the deformation selects a single symmetry-allowed resonant channel.

A first resonant normal form is obtained by averaging over the fast angle $\chi$ while retaining the $\cos(2\psi)$ term,
\begin{equation}
H_{\rm res}(I_+,I_-,\psi)= \omega_+ I_+ + \omega_- I_- + \lambda Z(I_+,I_-) + \lambda V(I_+,I_-)\cos(2\psi)\,,
\end{equation}
where we have defined the combinations $\omega_\pm=\omega_1\pm\omega_2$. The resulting dynamics is that of a one-degree-of-freedom resonant subsystem in $(I_-,\psi)$ with pendulum-like behaviour for $\psi$. Transport transverse to invariant tori is confined to a narrow resonant layer near $\omega_-\simeq0$; away from this region the oscillatory term averages out and KAM tori persist for sufficiently small $\lambda$.

In the Koopman-Krylov pipeline we seed Krylov space with the observables $g_Q$ and $g_{\rm mix}$ used in sec.~\ref{sec:T11}. In the restricted system, $g_Q$ in action-angle variables
\begin{equation}
g_Q \equiv Q = 2(\omega_1 I_1+\omega_2 I_2)\,.
\end{equation}
Thus $g_Q$ probes primarily the amplitudes: at fixed actions it is insensitive to the relative phase $\psi$ and varies only through slow action drift, which is suppressed away from the resonant layer.

By contrast, the mixing observable $g_{\rm mix}$, as defined in \eqref{def_T11_AdS_gmix}, depends directly on the relative phase. Substituting the action-angle parametrisation gives
\begin{equation}
g_{\rm mix} = \sqrt{I_1 I_2}\Bigl(\sqrt{\frac{\omega_2}{\omega_1}}-\sqrt{\frac{\omega_1}{\omega_2}}\Bigr)\sin\chi + \sqrt{I_1 I_2}\Bigl(\sqrt{\frac{\omega_2}{\omega_1}}+\sqrt{\frac{\omega_1}{\omega_2}}\Bigr)\sin\psi\,.
\end{equation}
Near the $1{:}1$ surface $\omega_1\simeq\omega_2$, the coefficient of $\sin\chi$ is suppressed as $O(|\omega_1-\omega_2|)$, while the coefficient of $\sin\psi$ remains $O(1)$.
Hence, precisely in the regime where the resonant term $\cos(2\psi)$ survives averaging and governs the slow dynamics, $g_{\rm mix}$ aligns with the slow angle and becomes a sensitive probe of the resonant channel.

This explains the observable hierarchy seen in sec.~\ref{sec:T11restricted} and in particular in fig.~\ref{fig:T11_restricted_W1_energy}. Indeed it follows that $g_Q$ mainly tracks slow action drift and responds weakly unless the dynamics strongly overlaps the resonant layer. On the other hand, $g_{\rm mix}$ directly probes the slow phase $\psi$ and responds sharply when the $1{:}1$ channel is active.

\paragraph{AdS-radial near-Penrose system.}
For the AdS-radial variant in sec.~\ref{sec:T11AdSrad}, the phase space is enlarged by an additional radial degree of freedom.
In contrast to the purely internal truncation, the deformation preserves the separable structure of the Hamiltonian: the radial motion remains dynamically decoupled from the internal oscillator sector, and the system admits a complete set of independent conserved quantities for all values of $\kappa$.
In particular, no small-denominator condition of the form $k_1\omega_1+k_2\omega_2\simeq0$ is induced by the deformation, and the interaction does not generate an effective resonant term such as $\cos(2(\phi_1-\phi_2))$ that would activate internal mode coupling.

The chosen observables remain $g_Q$ and $g_{\rm mix}$ as defined in eqs.\eqref{def_T11_AdS_gQ} and \eqref{def_T11_AdS_gmix}. Here $g_Q$ includes the AdS radial kinetic contribution, while $g_{\rm mix}$ continues to probe the internal relative phase. Since no deformation-induced internal resonance is present, the resonant normal-form mechanism described above does not operate, and one expects only mild deformation dependence in the Koopman spectral transport.

\bibliographystyle{JHEP}
\bibliography{biblio}
\end{document}